\documentclass{IEEEtran}

\usepackage{generic}
\usepackage{cite}
\usepackage{amsmath,amssymb,amsfonts}
\usepackage{algorithmic}
\usepackage{graphicx}
\usepackage{algorithm,algorithmic}
\usepackage{hyperref}
\hypersetup{hidelinks=true}
\usepackage{textcomp}

\usepackage{outline}
\usepackage{graphicx}
\usepackage{times}
\usepackage{bm} 
\usepackage{url}
\usepackage{epstopdf}
\usepackage{epsfig}
\usepackage{nicefrac}       
\usepackage{booktabs}       
\usepackage{graphicx}
\usepackage{multirow}
\usepackage{accents}
\usepackage{latexsym}
\usepackage{eso-pic}
\usepackage{tabu}
\usepackage{soul}
\usepackage{soulpos}

\usepackage{tikz}
\usepackage{colortbl}

\newcommand{\add}[1] {\textcolor{black}{#1}} 

\newcommand{\F}{\bm{F}}

\newcommand{\Z}{\bm{Z}}
\newcommand{\x}{\bm{x}}

\newcommand{\y}{\bm{y}}
\newcommand{\z}{\bm{z}}

\newcommand{\1}{\bm{1}}

\newcommand{\Ib}{{\bm I}}

\newcommand{\bb}{{\bm b}}

\newcommand{\db}{{\bm d}}

\newcommand{\jb}{{\bm j}}

\newcommand{\s}{{\bm s}}

\newcommand{\ob}{{\bm o}}

\newcommand{\Rc}{\mathcal{R}}

\newcommand{\Ac}{\mathcal{A}}

\newcommand{\Cc}{\mathcal{C}}

\newcommand{\Nc}{\mathcal{N}}

\newcommand{\Tc}{\mathcal{T}}
\newcommand{\Qc}{\mathcal{Q}}

\DeclareMathOperator*{\argmin}{arg\,min}

\newcommand{\Rd}{{\mathbb R}}

\newcommand{\epsilonb}{{\boldsymbol{\epsilon}}}

\newcommand{\Ed}{{{\mathbb E}}}

\definecolor{BrickRed}{rgb}{0.6,0,0}
\definecolor{RoyalBlue}{rgb}{0,0,0.8}
\definecolor{Tdgreen}{rgb}{0,0.4,0.7}
\definecolor{pinegreen}{rgb}{0.0, 0.47, 0.44}
\definecolor{cornellred}{rgb}{0.7, 0.11, 0.11}
\definecolor{cadmiumgreen}{rgb}{0.0, 0.42, 0.24}
\definecolor{spirodiscoball}{rgb}{0.06, 0.75, 0.99}

\definecolor{mylightblue}{rgb}{0.85, 0.90, 0.94}
\sethlcolor{mylightblue}
\definecolor{mylightred}{rgb}{0.90, 0.85, 0.94}

\def\code#1{\texttt{#1}}

\def\BibTeX{{\rm B\kern-.05em{\sc i\kern-.025em b}\kern-.08em
    T\kern-.1667em\lower.7ex\hbox{E}\kern-.125emX}}
\begin{document}
\title{Fundus Image Enhancement Through \\Direct Diffusion Bridges}
\author{Sehui Kim$^*$, Hyungjin Chung$^*$, Se Hie Park, Eui-Sang Chung, \\Kayoung Yi$^\dagger$, and Jong Chul Ye$^\dagger$, \IEEEmembership{Fellow, IEEE}
\thanks{
\add{
This work was supported by the National Research Foundation (NRF) of Korea under Grant RS-2024-00336454 and RS-2023-00262527, by the Korea Medical Device Development Fund grant funded by the Korean government (the Ministry of Science and ICT, the Ministry of Trade, Industry and Energy, the Ministry of Health \& Welfare, the Ministry of Food and Drug Safety, Project Number: 1711137899, KMDF\_PR\_20200901\_0015), by the MSIT(Ministry of Science and ICT), Korea, under the ITRC(Information Technology Research Center) support program(IITP-2024-2020-0-01461) supervised by the IITP(Institute for Information \& communications Technology Planning \& Evaluation), and by the Institute of Information \& communications Technology Planning \& Evaluation (IITP) grant funded by the Korea government(MSIT) (No.RS-2021-II212068, Artificial Intelligence Innovation Hub).
}
}
\thanks{$^*$: Equal contribution, $^\dagger$: corresponding authors.}
\thanks{Sehui Kim and Jong Chul Ye are with the Kim Jae Chul Graduate School of AI, KAIST, Daejeon, South Korea 34141. (e-mail: \{sehui.kim, jong.ye\}@kaist.ac.kr)}
\thanks{Hyungjin Chung is with the Bio \& Brain Engineering, KAIST, Daejeon, South Korea 34141. (e-mail: hj.chung@kaist.ac.kr)}
\thanks{Se Hie Park and Kayoung Yi are with Dept. of Opthalmology, Kangnam Sacred Heart Hospital, Hallym University School of Medicine, Seoul, South Korea. (e-mail: kayoungyi@yahoo.co.kr) Eui-Sang Chung is with the SNU Seoul Eye Clinic, Seoul, South Korea.}
}
\maketitle

\begin{abstract}
We propose FD3, a fundus image enhancement method based on direct diffusion bridges, which can cope with a wide range of complex degradations, including haze, blur, noise, and shadow.
We first propose a synthetic forward model through a human feedback loop with board-certified ophthalmologists for maximal quality improvement of low-quality in-vivo images.
Using the proposed forward model, we train a robust and flexible diffusion-based image enhancement network that is highly effective as a stand-alone method, unlike previous diffusion model-based approaches which act only as a refiner on top of pre-trained models.
Through extensive experiments, we show that FD3 establishes \add{superior quality} not only on synthetic degradations but also on in vivo studies with low-quality fundus photos taken from patients with cataracts or small pupils. \add{To promote further research in this area, we open-source all our code and data used for this research at \url{https://github.com/heeheee888/FD3}.}
\end{abstract}


\begin{figure*}[!thb]
\centerline{\includegraphics[width=\textwidth]{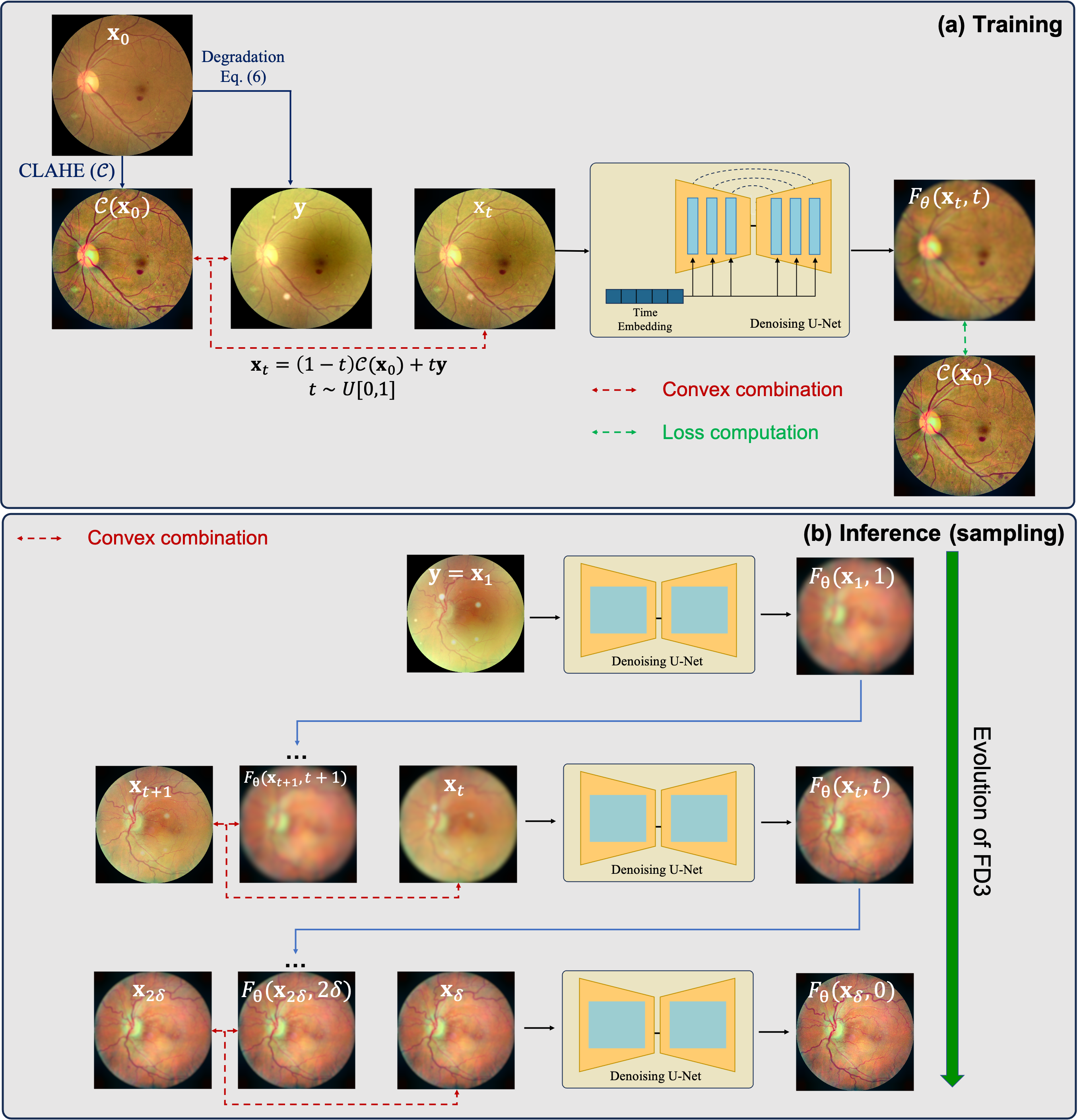}}
\caption{{(a) Training of FD3. CLAHE-applied high-quality images $\Cc(\x_0)$ are used as pseudo-ground-truth. $\x_t$ are randomly sampled to be convex combinations between $\Cc(\x_0)$ and the measurement $\y$. The neural network $F_\theta$ is trained to map any $\x_t$ to be close to $\Cc(\x_0)$. (b) Inference (sampling) of FD3. Trained neural network $F_\theta$ refines the posterior mean by following \eqref{eq:inference_fd3}, and directly starting from $\y = \x_1$. At every timestep, an approximate posterior mean $\Ed[\x_0|\x_t]$ is produced as a direct output of the neural network $F_\theta$.}}
\label{fig:fd3}
\end{figure*}

\section{Introduction}
\label{sec:intro}
\IEEEPARstart{F}{undus} photography is a crucial diagnostic tool used in ophthalmology to capture detailed images of the retina, such as the optic disc, macula, and blood vessels. These images, known as fundus photographs or fundus images, play a significant role in the detection, diagnosis, and monitoring of various eye conditions and systemic diseases that manifest in the eye~\cite{silva2015diabetic,issa2013macular,niemeijer2005automatic,wong2008natural}.
Unfortunately, the quality of the fundus photos is often hampered by various reasons, one of the most prominent being the media opacity from e.g. cataracts, and artifacts in the periphery arising from small pupils. 
Specifically, in a study involving more than 5,000 patients, more than 10\% of the photos were {\em inadequate} to be used for diagnosis~\cite{philip2005impact}.
Moreover, these difficulties were also pointed out in several other clinical studies, often leading to the need to discard the degraded photos completely from the study~\cite{wong2001retinal}.

In order to mitigate the degradations, the seminal work of \cite{peli1989restoration} designed the forward imaging model that focuses on the media opacity, along with simple solutions to the inverse problem. Notably, the devised forward model is, under simplifying assumptions, identical to the natural image {\em haze} forward model~\cite{narasimhan2002vision,fattal2008single}. Over the years, several studies have been conducted focusing on photos of the hazy fundus. Earlier works were mainly focused on the global characteristics of the image, i.e. histogram, and modifying them to enhance the visibility by means of, e.g. histogram equalization. One of the most effective algorithms along this line are the variants utilizing contrast limited adaptive histogram equalization (CLAHE)~\cite{setiawan2013color,jintasuttisak2014color,zhou2017color}. These methods are useful for enhancing the visibility of the internal structure such as vessels, but often cause unnatural shifts in color distribution and can only consider the existing global characteristics without prior knowledge.

More recently, deep learning-based approaches based on data have become dominant. However, there are several distinct caveats specifically for the fundus image enhancement problem that complicate the direct adoption of well-established supervised deep learning methods. One is that there is no standard consensus on the forward imaging model of the problem. For instance, the haze model introduced by~\cite{peli1989restoration} only models the effect of the internal turbid medium. A more complicated model that attempts to encapsulate several of the external effects such as motion blur or halo artifacts can potentially be used~\cite{shen2020modeling}, but it is still unknown whether such a process truly approximates the imaging system. The other complication is that it is extremely hard to collect datasets that are paired, i.e. aligned. The best way to construct such a dataset is by taking the fundus photos that were taken before and after the cataract surgery, which would still not guarantee perfect alignment as there is high variance stemming from other factors, e.g., inter-operator variance. 

Taking into account the difficulties, deep learning methods can be largely classified into three categories: 1) Adopting the naive forward model of~\cite{peli1989restoration} and aiming to solve the inverse problem by supervised, or unsupervised learning~\cite{qayyum2020single,qayyum2022single}, 2) {\em learning} forward imaging through GAN training, and training in a supervised fashion from the simulated paired dataset~\cite{luo2020dehaze}, 3) Carefully designing a realistic degradation model to simulate the dataset and performing supervised training on the dataset~\cite{shen2020modeling}. To this end, we propose a method, named \textbf{F}undus \textbf{D}egradation enhancement through \textbf{D}irect \textbf{D}iffusion (FD3). FD3 follows along the line of 3), but introduces several key contributions:

\begin{itemize}
    \item We propose the first diffusion model-based fundus image enhancement scheme that achieves \add{superior} quality as a standalone method and does not rely on other pre-trained enhancement modules \add{(See Fig.~\ref{fig:fd3} for the overview of the proposed method)}
    \item \add{We elucidate the advantage of direct diffusion bridges over standard denoising diffusion models by revealing their key similarities and differences, and thereby adopt the former approach suited specifically for fundus photo enhancement}
    \item We provide a fix to the forward model used for simulation, which is particularly strong for enhancing the quality of {\em real} degradations
    \item We perform an extensive evaluation using both the simulated forward models and in-vivo data, with standard quantitative metrics as well as evaluations from board-certified ophthalmologists with ample experience.
    \item \add{To promote further research in the area, we open-source all our code and data used in this study at \url{https://github.com/heeheee888/FD3}}
\end{itemize}

\section{Background}
\label{sec:background}

\subsection{Imaging forward model of fundus photos}
\label{sec:imaging_forward_model}

The degradation process of the fundus photo is complex and highly stochastic due to the interoperator variance. Here, we review some of the widely used forward models in the literature, which will be useful for defining the forward model used throughout this work. \cite{peli1989restoration} proposed a forward process similar to a hazing process for natural images~\cite{he2010single}, which reads
\begin{align}
\label{eq:dehazing_forward_model}
    \y = \jb \odot \x + a (\1 - \jb), \quad \x \in \Rd^n,\,\y \in \Rd^n,
\end{align}
where $\jb \in \Rd^n$ acts as attenuation, $\odot$ denotes element-wise product, $\1$ refers to the vector of 1s, and $a \in \Rd$ is the atmospheric light assumed to be constant across the whole image. For the case of dehazing, $\jb$ is related to the {\em depth} $\db$ of the scene
\begin{align}
    \jb = \exp (- \delta \db), \quad \db \in \Rd^n
\end{align}
where $\delta$ is the scattering coefficient. This forward has been used in e.g.\cite{qayyum2020single,qayyum2022single} together with the use of deep image prior (DIP)~\cite{ulyanov2018deep} for the enhancement of fundus photos.

Nevertheless, it was pointed out that \eqref{eq:dehazing_forward_model} is too simplistic to fully capture the degradation process of the fundus photos. Consequently, \cite{shen2020modeling} proposed three distinct components of the forward process: light transmission distortion, blur, and retinal artifacts. 
Light transmission distortion can be mathematically written as
\begin{align}
\label{eq:light_transmission_distortion}
    \Tc(\x) := {\rm clip}(\alpha(B_{\phi_l}\bb + \x) + \beta; \gamma),
\end{align}
where $\alpha$ is the contrast factor, $\beta$ is the brightness offset, ${\rm clip(\cdot; \gamma)}$ is the clipping operator at the value $\gamma$, and $B_{\phi_l}$ is the Gaussian blur operator with the parameter $\phi_l$. 
The illumination bias $\bb \in \Rd^n$ is a vector with non-zero values that represent over/under-illumination in a disc-shaped region of the image. The blurring is defined as
\begin{align}
\label{eq:blurring}
    \Qc(\x) := B_{\phi_b} \x + \bm{\eta}, \quad \bm{\eta} \sim \Nc(\bm{0}, \sigma_y^2\Ib),
\end{align}
\add{where $\Nc$ denotes the Gaussian distribution,} $\phi_b$ is the parameter for the blurring Gaussian kernel, and $\bm{\eta}$ denotes additive white Gaussian noise with variance $\sigma_y^2\Ib$. Finally, the retinal artifact is defined as
\begin{align}
\label{eq:retinal_artifact}
    \Rc(\x) := \x + \sum_{i=1}^N B_{\phi_o}\ob_i,
\end{align}
where $\ob_i \in \Rd^n$ are vectors with non-zero values on the disc-shaped region of the image, similar to but smaller than $\bb$. In \cite{shen2020modeling}, the authors use \eqref{eq:light_transmission_distortion},\eqref{eq:blurring},\eqref{eq:retinal_artifact} in conjunction
\begin{align}
\label{eq:forward_conjunction}
    \y = \Ac(\x) := \Rc \circ \Qc \circ \Tc(\x),
\end{align}
with parameters of $\Rc,\Qc,\Tc$ sampled randomly to simulate the forward process for training supervised neural networks, where $\circ$ denotes function composition.

\subsection{Diffusion models and inverse problems}
\label{sec:diffusion_models}

Diffusion models~\cite{sohl2015deep,ho2020denoising,song2020score} are a class of generative models that learn to reverse the forward Gaussian noising process. The process is usually defined with a time horizon $t \in [0, 1]$ with $p_0(\x_0) := p_{\rm data}(\x_0)$ and $p_1(\x_1) \approx \Nc(0, \Ib)$. 
A typical diffusion model takes a Gaussian perturbation kernel through time $t$, which can be defined as $p(\x_t|\x_0) = \Nc(\x_t; s_t\x_0, s_t\sigma_t^2\Ib)$. Several choices can be made to ensure $p_1(\x_1) \approx \Nc(0, \Ib)$, e.g. variance preserving (VP), variance exploding (VE) formulation of \cite{song2020score}, or a more simplified form of taking $s_t = 1, \sigma_t = t$ as in \cite{karras2022elucidating}. Under this latter choice, the probability-flow ordinary differential equation (PF-ODE)~\cite{song2020score} of generating noise from data can be represented as
\begin{align}
\label{eq:pf_ode}
    d\x_t = -t\nabla_{\x_t}\log p(\x_t)\,dt = \frac{\x_t - \Ed[\x_0|\x_t]}{t},
\end{align}
where the transition between the score function $\nabla_{\x_t} \log p(\x_t)$ and the posterior mean $\Ed[\x_0|\x_t]$ is given by the Tweedie's formula~\cite{efron2011tweedie}, which states $\Ed[\x_0|\x_t] = \x_t + t^2\nabla_{\x_t}\log p(\x_t)$. In practice, one can estimate the score function by using the denoising score matching (DSM) loss~\cite{vincent2011connection}
\begin{align}
\label{eq:dsm}
    \min_\theta \Ed_{\x_t,\x_0,\epsilonb}\left[ \|\s_\theta(\x_t,t) - \nabla_{\x_t} \log p(\x_t|\x_0)\|_2^2 \right].
\end{align}
Once the neural network $\s_{\theta^*}$ is trained, it can be used as a plug-in estimate to \eqref{eq:pf_ode}. Consequently, the reverse SDE starts with sampling a random Gaussian noise vector $\x_1 \sim \Nc(0, \Ib)$, and solving \eqref{eq:pf_ode} with a numerical method, which amounts to sampling from $p_\theta(\x_0)$.

It was shown that one can solve various inverse problems through the pre-trained diffusion model~\cite{kadkhodaie2021stochastic,kawar2022denoising,chung2023diffusion} simply by modifying the reverse diffusion of \eqref{eq:pf_ode}, replacing $\nabla_{\x_t} \log p(\x_t)$ with $\nabla_{\x_t} \log p(\x_t|\y)$. A limitation of these approaches is that one has to know the exact forward model that generated the measurement, which is a condition that is often unmet in real-world problems. Subsequently, non-blind inverse problem solvers were extended to blind inverse problems in \cite{chung2023parallel,murata2023gibbsddrm}, targetting applications such as blind deblurring. While these methods are useful for problems where we know the functional form of the forward model a priori (e.g. convolution with a kernel), they are hard to apply when the imaging model is inaccurate or ambiguous. Remarkably, this is the case for fundus photography enhancement, where the forward model is highly stochastic and relatively inaccurate. We empirically show that this is indeed the case in Sec.~\ref{sec:blinddps_comparison}.

\subsection{Direct Diffusion Bridge}
\label{sec:ddb}

\add{Chung {\em et al.} \cite{chung2023direct} unified the seemingly different approaches of InDI~\cite{delbracio2023inversion} and I2SB~\cite{liu2023i2sb}} into a single framework called direct diffusion bridge (DDB). Namely, given a paired tuple $(\x, \y) \sim p(\x, \y)$ and $\x_0 := \x \sim p(\x)$, $\x_1 := \y \sim p(\y|\x)$, DDB defines the diffusion process $p(\x_t|\x,\y)$ as
\begin{align}
\label{eq:ddb_diffusion_process}
    \x_t = (1 - \alpha_t)\x_0 + \alpha_t\x_1 + \sigma_t\z,\, \z \sim \Nc(0,\Ib)
\end{align}
where $\z \sim \Nc(0, \Ib)$, and $\{\alpha_t, \sigma_t\}_{t=0}^1 \in [0, 1]$ are the signal/noise schedules that govern the process. A neural network $F_\theta$ is trained to estimate the posterior mean
\begin{align}
\label{eq:ddb_training}
    \theta^* = \argmin_\theta \Ed_{(\x,\y) \sim p(\x,\y),t \sim p(t)}[\|F_\theta(\x_t) - \x\|_2^2],
\end{align}
such that $\F_{\theta^*}(\x_t) \approx \Ed[\x|\x_t], \forall t$. The inference distribution $p(\x_s|\x_0,\x_t)$ is defined analogous to \add{denoising diffusion probabilistic models}(DDPM)~\cite{ho2020denoising}, which can be written with reparametrization trick as
\begin{align}
     \x_s = (1 - \alpha_{s|t}^2)\x_0 + \alpha_{s|t}^2\x_t + \sigma_{s|t}\z,\,\z \sim \Nc(0, \Ib),
\end{align}
with $\alpha_{s|t}, \sigma_{s|t}$ chosen so that the variance condition of the marginal distribution is met.

\section{Main Contribution: The FD3 Algorithm}
\label{sec:main}

\begin{figure}
\centerline{\includegraphics [width=8cm]{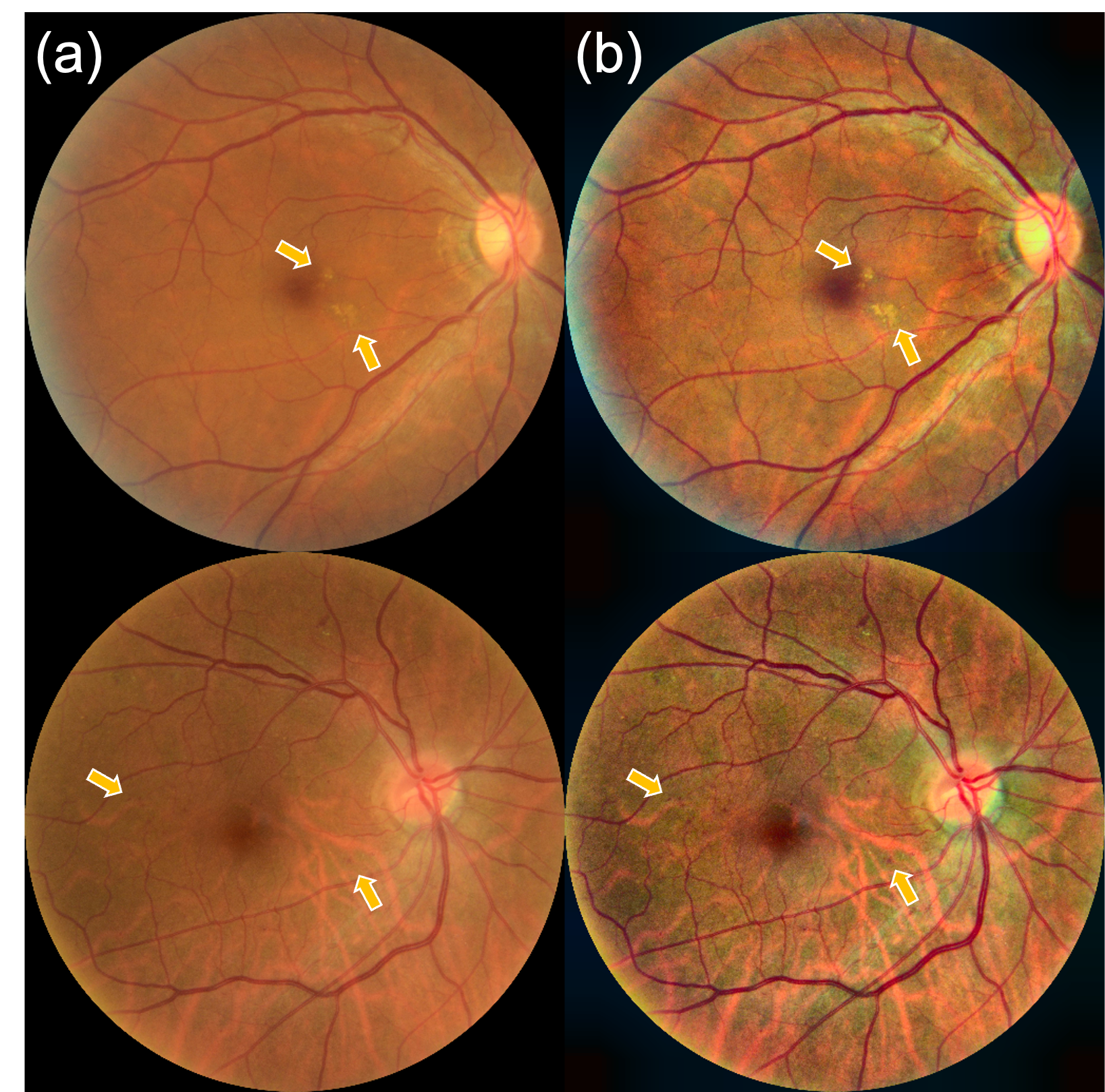}}
\caption{(a) Before, (b) After applying CLAHE to ``ground-truth'' images. 1$^{\rm st}$ row: drusen marked with yellow arrows. 2$^{\rm nd}$ row: hemorrhage and microaneurysm marked with yellow arrows.}
\label{fig:clahe_example}
\end{figure}

\subsection{Synthethic forward model}

One of the most important factors in fundus photo enhancement is the visibility of the internal structure.
This has led to the wide popularity of the usage of Contrast Limited Adaptive Histogram Equalization (CLAHE)~\cite{setiawan2013color,jintasuttisak2014color,zhou2017color}, which alters the global characteristics of the color image histogram. A clear example of the enhancement in the visibility of the vessels can be seen in Fig.~\ref{fig:clahe_example}.
However, the forward model of \cite{shen2020modeling} defined in \eqref{eq:forward_conjunction} only considers the illumination ($\Tc(\cdot)$, \eqref{eq:light_transmission_distortion}), blur ($\Qc(\cdot)$, \eqref{eq:blurring}), and artifacts ($\Rc(\cdot)$, \eqref{eq:retinal_artifact}), disregarding the global characteristics.
To overcome this drawback, we construct a process that is defined by
\begin{align}
\label{eq:proposed_forward}
    \y = \Ac \circ \Cc^{-1}(\x) =: \tilde\Ac(\x),
\end{align}
where $\Cc(\cdot)$ represents the CLAHE operation, and $\Cc^{-1}$ its inverse. This is equivalent to considering $\Cc(\x)$ as the ``ground truth'' in \eqref{eq:forward_conjunction}. The motivation for constructing \eqref{eq:proposed_forward} is to utilize a ground truth image of the highest quality. In other words, we are assuming that even the ``high-quality'' images established in the widely-used datasets such as EyeQ~\cite{Fu_2019} are not optimal in perceptual quality, such that applying $\Cc(\cdot)$ results in a better representation of the desired image. 

To show that this is indeed the case, two board-certified ophthalmologists conducted a qualitative analysis of the two types of images, evaluating the ``goodness'' of the images. Overall, the quality of the images after applying CLAHE was chosen to be of relatively {\em better} quality than the ground truth images before CLAHE. Specifically, in the first row of Fig.~\ref{fig:clahe_example}, drusen were more visible. In the second row, diabetic retinal changes, represented by hemorrhage and micro-aneurysm, were more clearly seen, enabling easier anomaly detection from the photo.

One may question the possibility of using the forward model defined in \eqref{eq:forward_conjunction} as is, and using CLAHE as a post-processing step. We show that this is suboptimal in Section~\ref{sec:in_vivo_study}, where we clearly see that we yield results that are sharper and with enhanced visibility. This can be attributed to the generalizability of the neural network, which leads to a better solution through amortization.

\begin{figure}[!t]
\centerline{\includegraphics [width=0.8\columnwidth]{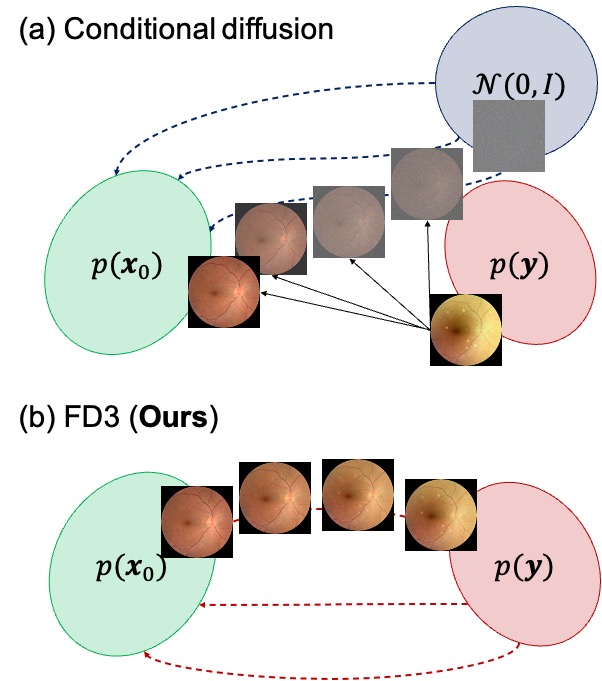}}
\caption{{Schematic illustration of (a) standard conditional diffusion, and (b) FD3. FD3 is capable of following a direct and smoother trajectory from $p(\y)$ to $p(\x_0)$, compared to the standard diffusion path, which involves starting the process from irrelevant Gaussian noise.}}
\label{fig:concept}
\end{figure}

\subsection{Direct bridge for fundus enhancement}
\label{sec:fd3}

Our goal is to construct a direct diffusion bridge that is able to revert the process of \eqref{eq:proposed_forward}. To achieve this goal, we construct a DDB by choosing the parameter $\alpha_t = t, \sigma_t = 0$. Of note, such choice has been consistently shown that such choice is effective in a wide variety of works, including \cite{liu2023flow,delbracio2023inversion}. Now, in order to leverage our proposed forward model in \eqref{eq:proposed_forward}, given a tuple $(\x_0:= \Cc(\x), \y:= \Ac(\x)) \sim p(\x,\y)$ with $\y = \x_1$, the diffusion process is then simply defined as a convex combination of the two components
\begin{align}
\label{eq:diffusion_fd3}
    \x_t = (1 - t)\x_0 + t\x_1, \quad {\mbox {\rm with}} \quad t \in [0, 1].
\end{align}
Under the diffusion process in \eqref{eq:diffusion_fd3}, we train a neural network to estimate the posterior mean $\Ed[\x_0|\x_t]$ with the following objective
\begin{align}
\label{eq:training_fd3}
    \theta^* = \argmin_{\theta}\Ed_{\x,\y \sim p(\x, \y), t \sim p(t)}\|F_\theta(\x_t,t) - \x\|_2^2,
\end{align}
where we take $p(t) \sim U[0, 1]$, and use a uniformly weighted loss to train the network. Once the network is trained with \eqref{eq:training_fd3}, we can perform inference by iteratively running
\begin{align}
\label{eq:inference_fd3}
    \x_s = (1 - \frac{s}{t})F_{\theta^*}(\x_t,t) + \frac{s}{t}\x_t, \quad s < t,
\end{align}
starting from $t = 1$, and taking uniform incremental steps. 
See Fig.~\ref{fig:fd3} for the schematic illustration of the inference process.

Here, recall that $F_{\theta^*}(\x_t,t) \approx \Ed[\x_0|\x_t]$ due to the design of the loss function in \eqref{eq:training_fd3}. This is a useful fact when performing enhancement through \eqref{eq:inference_fd3}, as at every step, one would always be estimating the posterior mean, or to put it another way, the most probable restoration on average, given the current estimate $\x_t$. Hence, one-step inference of taking $s = 0, t = 1$ would give us $\Ed[\x|\y]$ directly, minimizing the pixel-wise error, or the so-called distortion~\cite{blau2018perception}. However, simply resorting to such one-step inference would yield results that are typically blurry, due to the regression-to-the-mean effect, studied extensively in, e.g.,~\cite{blau2018perception,delbracio2023inversion,chung2023direct}. Instead, taking multiple steps of \eqref{eq:inference_fd3} will iteratively refine the posterior mean, such that the ending result will have better perceptual quality. In fact, this is closely related to the fact that (see further discussion in \cite{delbracio2023inversion})
\begin{align}
    \Ed[\x_s|\x_t] = \left(1 - \frac{s}{t}\right)\Ed[\x_0|\x_t] + \frac{s}{t}\x_t.
\end{align}
Thus, the inference in \eqref{eq:inference_fd3} would lead to taking small-step minimum mean-squared error (MMSE) estimates.

\noindent
\textbf{\add{Desirable path}.~}
Setting $s \rightarrow t$ for \eqref{eq:training_fd3} leads to an ordinary differential equation (ODE).
\begin{align}
\label{eq:fd3_ode}
    d\x_t = \frac{\x_t - F_{\theta^*}(\x_t)}{t}\,dt \approx \frac{\x_t - \Ed[\x_0|\x_t]}{t}.
\end{align}
Due to the design choice of FD3, $\x_1 = \y$, hence running \eqref{eq:fd3_ode} would lead to a smooth bridge that starts from our measurement $\y$ that is gradually transitioned to $\x_0$. On the other hand, consider running posterior sampling with the standard diffusion model by conditioning \eqref{eq:pf_ode} with $\y$
\begin{align}
\label{eq:pf_ode_posterior}
    d\x_t = -t\nabla_{\x_t}\log_p(\x_t|\y)\,dt = \frac{\x_t - \Ed[\x_0|\x_t,\y]}{t}.
\end{align}
Here, we notice the surprising similarity between \eqref{eq:pf_ode},\eqref{eq:pf_ode_posterior} and \eqref{eq:fd3_ode}. The only difference in the two different types of ODEs is that for the diffusion PF-ODE, one starts sampling from $\x_1 \sim p(\x_1) = \Nc(0, \Ib)$, a standard Gaussian noise independent of the given problem, and for FD3, one can start sampling from $\y$, the measurement that we would like to enhance. Hence, using the standard diffusion path yields a considerably more complex inference process, whereas, for FD3, we can use a much more direct, smoother path starting from $\y$. 
See Fig.~\ref{fig:concept} for an illustration of the difference between the two algorithms. 
As a consequence, FD3 requires much less compute (e.g. 5 NFE) as opposed to the heavy compute needed for standard conditional diffusion models (e.g. 1000 NFE).

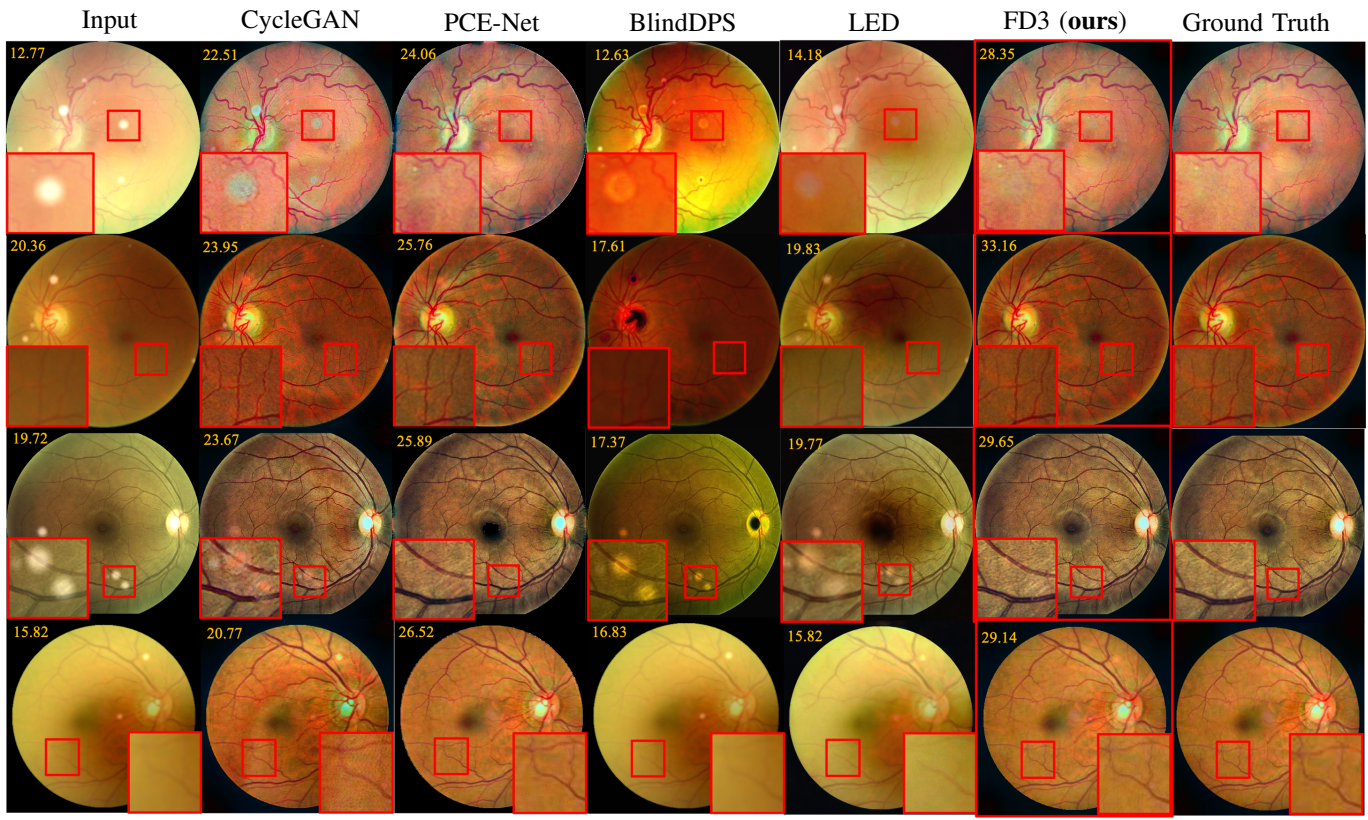
\begin{figure*}[!thb]
    \begin{tikzpicture}
        \node[anchor=south west,inner sep=0] (image) at (0,0) {\includegraphics[width=\textwidth]{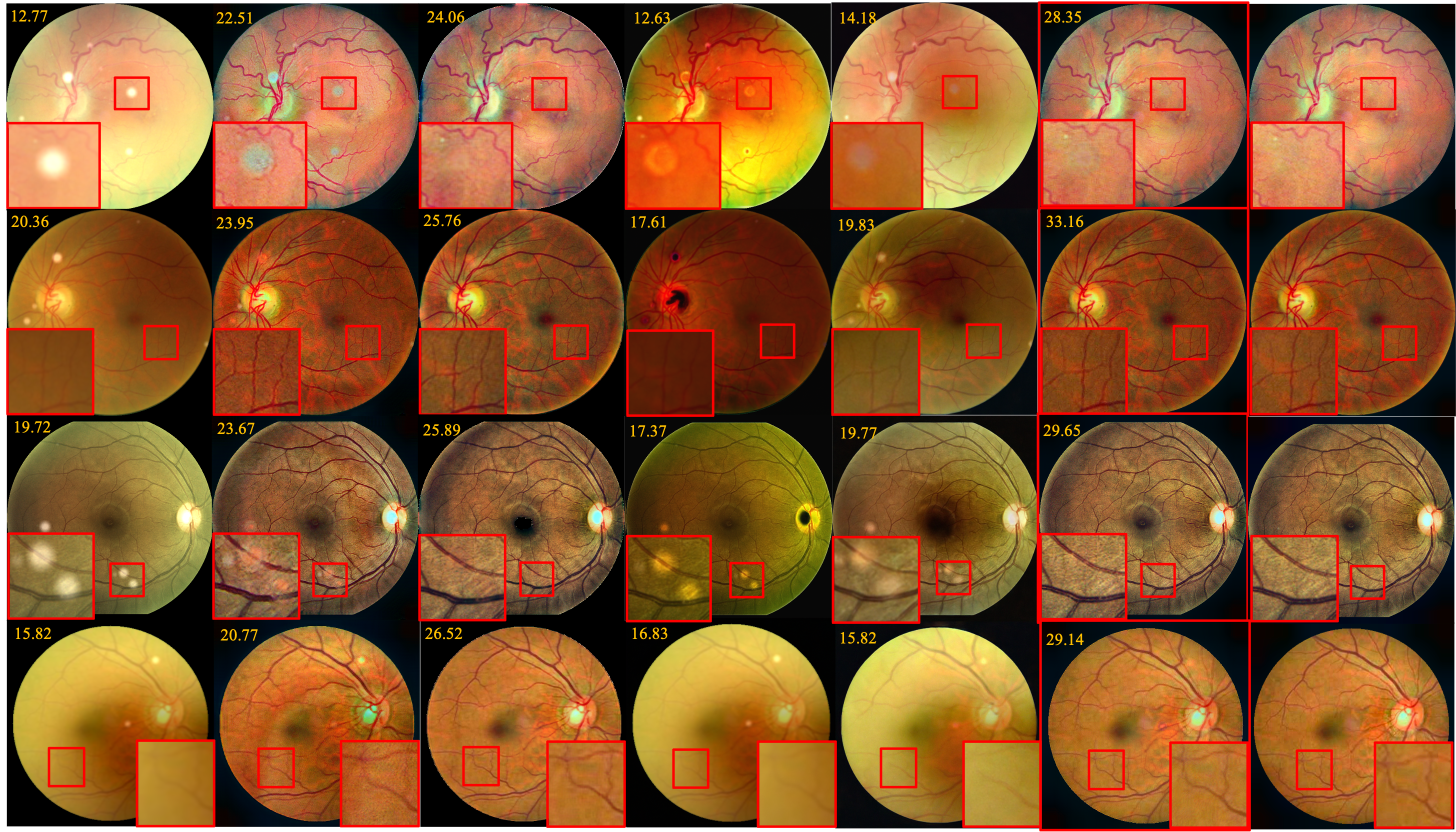}};
        
        \begin{scope}[x={(image.south east)},y={(image.north west)}]
            \node[text width=3cm, align=center] at (0.08,1.02) {Input};
            \node[text width=3cm, align=center] at (0.22,1.02) {CycleGAN};
            \node[text width=3cm, align=center] at (0.36,1.02) {PCE-Net};
            \node[text width=3cm, align=center] at (0.50,1.02) {BlindDPS};
            \node[text width=3cm, align=center] at (0.64,1.02) {LED};
            \node[text width=3cm, align=center] at (0.78,1.02) {FD3 (\textbf{ours})};
            \node[text width=3cm, align=center] at (0.92,1.02) {Ground Truth};
        \end{scope}
    \end{tikzpicture}
    \caption{\add{(\textbf{Simulation study}) Comparison of the image enhancement quality using our proposed forward model. From 1st column to 3rd column: EyeQ dataset, 4th column: FPE dataset, CycleGAN~\cite{zhu2017unpaired}, PCE-Net~\cite{Liu_2022}, BlindDPS~\cite{chung2023parallel}, LED~\cite{cheng2023learning}, FD3 (\textbf{Ours}), and ground truth. Yellow numbers in the top left corner: PSNR.}}
    \label{fig:results_simulation_main}
\end{figure*}

\begin{figure*}[!thb]
    \begin{tikzpicture}
        \node[anchor=south west,inner sep=0] (image) at (0,0) {\includegraphics[width=\textwidth]{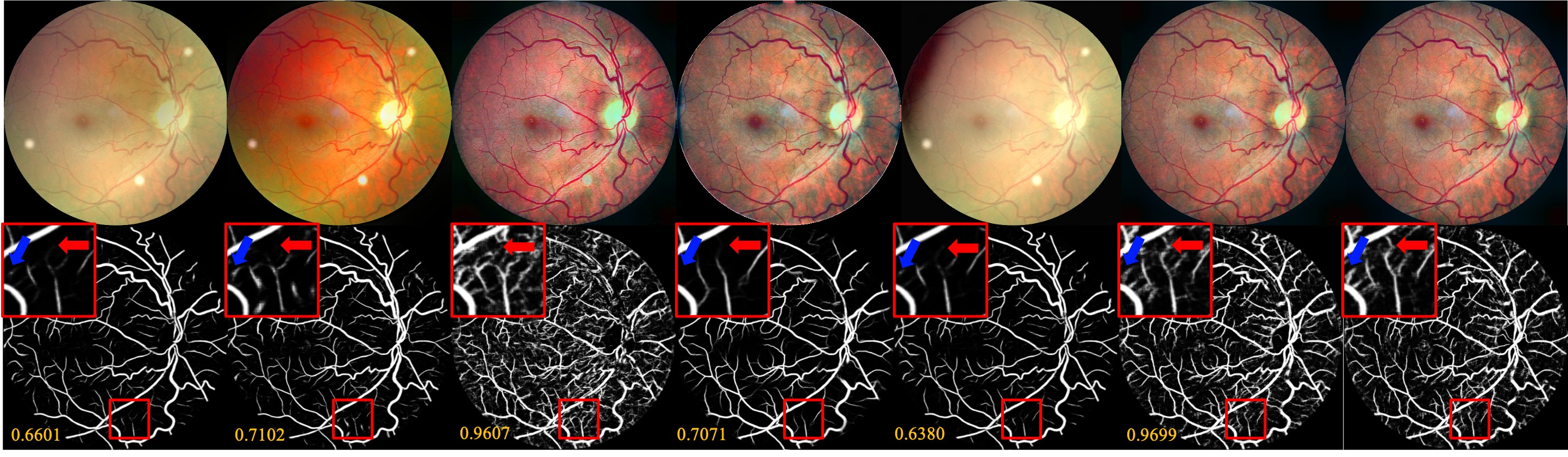}};
        
        \begin{scope}[x={(image.south east)},y={(image.north west)}]
            \node[text width=3cm, align=center] at (0.08,1.04) {Input};
            \node[text width=3cm, align=center] at (0.22,1.04) {DCP-BCP};
            \node[text width=3cm, align=center] at (0.36,1.04) {CycleGAN};
            \node[text width=3cm, align=center] at (0.50,1.04) {PCE-Net};
            \node[text width=3cm, align=center] at (0.64,1.04) {BlindDPS};
            \node[text width=3cm, align=center] at (0.78,1.04) {FD3 (\textbf{ours})};
            \node[text width=3cm, align=center] at (0.92,1.04) {Ground Truth};
        \end{scope}
    \end{tikzpicture}
    \caption{\add{{Downstream vessel segmentation performance evaluation using a pre-trained model Iter-Net~\cite{9093621}. Yellow numbers on the bottom left corner: IOU.}}}
    \label{fig:vessel}
\end{figure*}
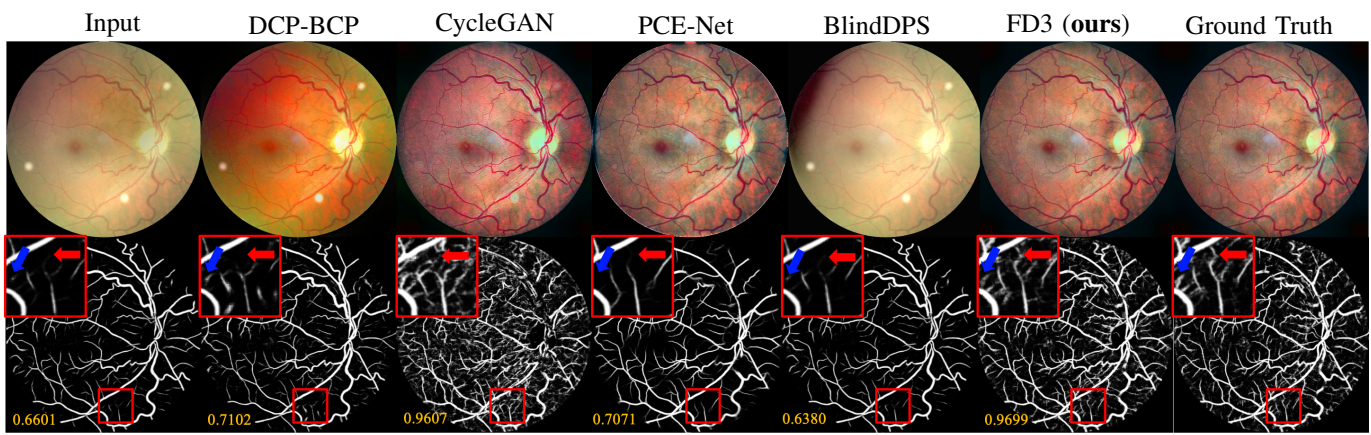

\begin{figure*}[!thb]
    \begin{tikzpicture}
        \node[anchor=south west,inner sep=0] (image) at (0,0) {\includegraphics[width=\textwidth]{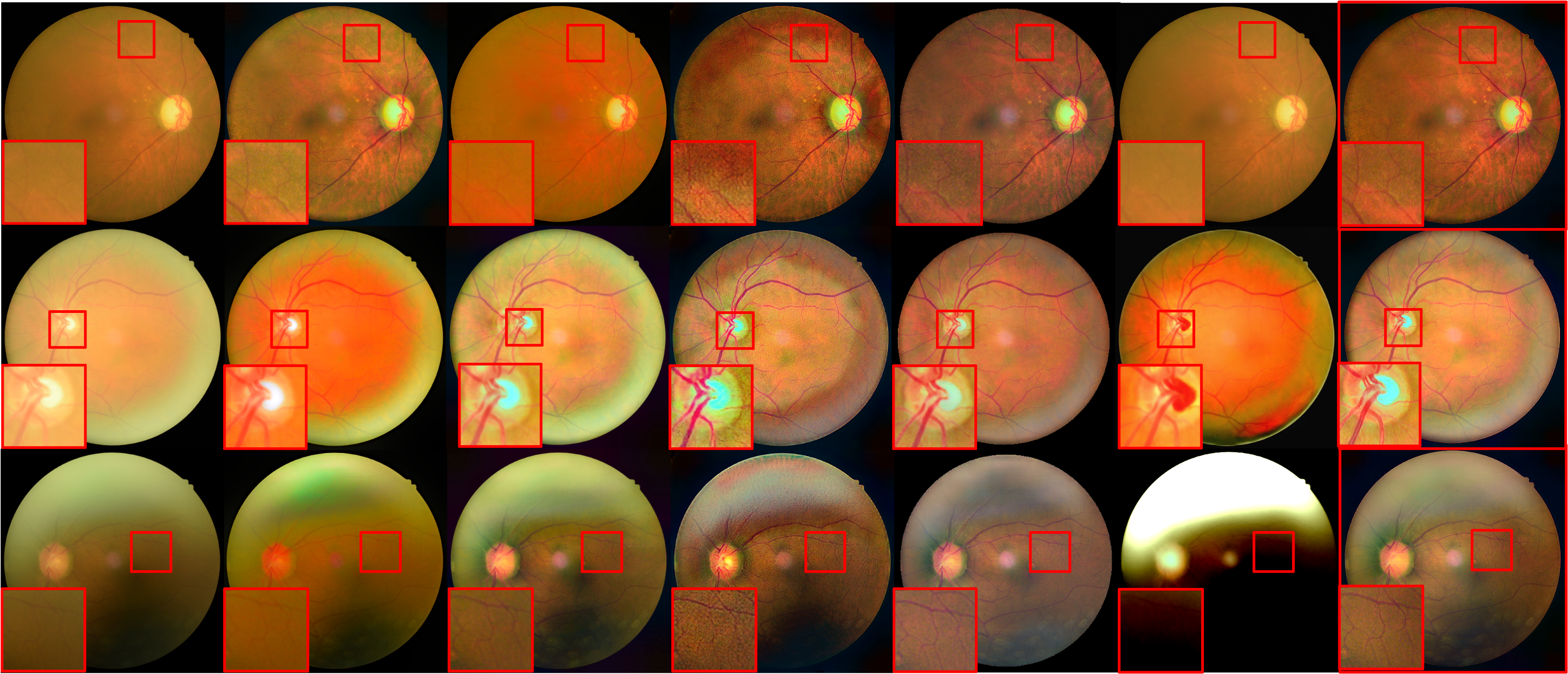}};
        
        \begin{scope}[x={(image.south east)},y={(image.north west)}]
            
            \node[text width=3cm, align=center] at (0.08,1.02) {Input};
            \node[text width=3cm, align=center] at (0.22,1.02) {DCP-BCP};
            \node[text width=3cm, align=center] at (0.36,1.02) {CLAHE};
            \node[text width=3cm, align=center] at (0.50,1.02) {CycleGAN};
            \node[text width=3cm, align=center] at (0.64,1.02) {SCR-Net};
            \node[text width=3cm, align=center] at (0.78,1.02) {BlindDPS};
            \node[text width=3cm, align=center] at (0.92,1.02) {FD3 (\textbf{ours})};
        \end{scope}
    \end{tikzpicture}
    \caption{\add{(\textbf{In-vivo study}) Comparison of the image enhancement quality using our proposed forward model. From 1st column to last column: bad quality image, DCP-BCP~\cite{he2010single}, CLAHE~\cite{setiawan2013color,jintasuttisak2014color,zhou2017color},  CycleGAN~\cite{zhu2017unpaired}, SCR-Net~\cite{li2022structureconsistent}, BlindDPS~\cite{chung2023parallel},FD3 (\textbf{Ours}).}}
    \label{fig:results_invivo_main}
\end{figure*}

\section{\add{Experimental settings}}
\label{sec:exp}

\noindent\textbf{Dataset.~}
We first perform a simulation study to quantitatively evaluate our method. Here, the EyeQ dataset ~\cite{Fu_2019} is used to verify the validity of our model. We only chose the fundus photos under the ``Good'' category, which consists of 16817 images in total. Among them, 15817 images were used for training, and the remaining 1000 were used for testing. 
For the simulated forward model, we choose two different types: 1) The first type of forward model exactly follows \cite{shen2020modeling}, expressed succinctly in \eqref{eq:forward_conjunction}; 2) The second type of forward model is the one that we devise, given by \eqref{eq:proposed_forward}, which includes applying CLAHE on top of light transmission disturbance, image blurring, and retinal artifacts.

Furthermore, we collected fundus photos from the Kangnam Sacred Heart Hospital, Hallym University School of Medicine, Seoul, South Korea (IRB approval number: 2022-10-026), that were obtained from 2000 to September 2022. The fundus photographs were taken by five skilled examiners using the KOWA Nonmyd 8S Fundus Camera (KOWA company, Japan). Among the 2,000 images, we chose 50 test images that were characterized as ``bad-quality'' due to one of the following reasons: media opacity, small pupil, or poor patient cooperation. For the rest of the 1,950 ``good-quality'' images, we removed the duplicates and those having sizes smaller than $256\times 256$. After the filtering, there are 1,152 images that are under the ``good-quality'' category.
We open-source all the images under the name \textbf{F}undus \textbf{P}hoto \textbf{E}nhancement (FPE) dataset to promote further research.

Using the FPE data, we conduct two types of additional studies. First, similar to the EyeQ experiment, we retrospectively degrade the ``good-quality'' images with our proposed synthetic forward model, to quantitatively evaluate the performance of the proposed method. Then, we perform a study on the 50 in-vivo low-quality images. All the images used in the experiments were resized and center-cropped to $[512,512]$ resolution.

\noindent\textbf{Training.~}
We used the model architecture based on \add{ablated diffusion model} (ADM)~\cite{dhariwal2021diffusion}. 
\add{
The details of the network architecture can be found in \url{https://github.com/heeheee888/FD3/blob/master/configs/optic.yml}. We use standard ResBlocks of ADM with the attention layer in only the coarsest resolution of the features and without any dropout.
}
Time embedding was randomly selected with a uniform distribution. The model was trained by the AdamW optimizer for 30 epochs with a learning rate of 0.0001, and a training batch size of 4.

\noindent\textbf{Inference.~}
For all experiments with FD3, we use 10 \add{neural function evaluation} (NFE) sampling with uniform discretization when iteratively applying \eqref{eq:inference_fd3}, i.e. we take $s = 0.9, 0.8, \dots, 0.0$, unless specified otherwise. Note that this is a design choice, which we explore further in Section.~\ref{sec:ablation_study}.

\noindent\textbf{Comparison methods.~}
To demonstrate the effectiveness of FD3, we compare against the representative baseline methods: CLAHE~\cite{setiawan2013color,jintasuttisak2014color,zhou2017color}, DCP-BCP~\cite{he2010single,wang2013automatic}, CycleGAN~\cite{zhu2017unpaired}, SCR-Net~\cite{li2022structureconsistent}, PCE-Net~\cite{liu2022pseudo}, \add{BlindDPS~\cite{chung2023parallel}, and LED~\cite{cheng2023learning}}.
Note that applying DCP or BCP a standalone method is one of the standard approaches for image enhancement. However, we find that applying the two methods sequentially resulted in superior performance in resolving the over/under-illumination of the imaging medium. Hence, we compare against this approach.
\add{
For implementing CLAHE, we used the \code{cv2.createCLAHE} function and found the best \code{clipLimit} and \code{tileGridSize} found through grid search. The parameters that we used throughout the experiments were set to \code{clipLimit}=2.0, \code{tileGridSize}=(8,8). The hyperparameters were chosen to preserve the original color fidelity and maintain the local contrast. When the tile size is reduced below (8,8), noise within local patches becomes excessively pronounced. Conversely, when the tile size exceeds ours, the method diminishes consideration of local contrast, resulting in images that deviate significantly from the originals in terms of overall color. Lowering the clip limit below 2.0 yields minimal disparity between the original and processed images. However, surpassing our clip limit leads to exaggerated noise, contrary to our intended outcome. Our choosing clip limit changed the contrast well, to the extent that it aided in distinguishing the microvascular and disease well.
}
CycleGAN~\cite{zhu2017unpaired} was trained for a total of 200 epochs. During the first 100 epochs, a learning rate was 0.0002 and was linearly decayed to zero over the last 100 epochs. \add{Both SCR-Net~\cite{li2022structureconsistent} and PCE-Net~\cite{liu2022pseudo} were also trained for 30 epochs with the learning rate 0.0002.} All three models employed a generator based on the ResNet with 9 blocks, to match the parameter count of the ADM model used for the proposed method.
\add{
For LED, we use the original implementation of LED in the following repository with default settings \url{https://github.com/QtacierP/LED}.
}

\noindent\textbf{Quantitative Evaluation Metric.~} 
We evaluated the results with peak signal-to-noise ratio (PSNR) to measure the distortion from the ground truth, Frechet inception distance (FID) to measure the perceptual quality, and the intersection over union (IOU) score of vessel segmentation to measure the downstream task performance. 

\add{
The PSNR metric between the ground truth $\x$ and its estimate $\hat\x$ is defined as
\begin{align}
\label{eq:psnr}
    {\rm PSNR}(\x, \hat\x) = 20 \log_{10} \left( \frac{MAX(\x)}{\sqrt{MSE(\x, \hat\x)}} \right),
\end{align}
where $MAX(\cdot)$ is the maximum pixel value of $\x$, and $MSE(\cdot, \cdot)$ computes the mean squared error between the two arguments.
To compute the FID, note that we first need to obtain the {\em distribution} of the features acquired from the \code{pool3} layer of the Inception network~\cite{szegedy2015going}
\begin{align}
    \z^i = f(\x^i), \quad \z^i \in \Rd^k,\, i = 1, \cdots, N,
\end{align}
where $\z^i$ is the feature vector of the $i^{\rm th}$ image obtained from the network $f$. After extracting the features, we fit the parameters by assuming that the feature vectors form a Gaussian distribution.
}
\add{
This is done separately for the feature vectors $\z^i$ acquired from the ground truth images to form $\Nc(\Z; \mu_{\Z}, \sigma_{\Z}^2 I)$
and the feature vectors $\hat\z^i$ acquired from the enhanced images to form $\Nc(\hat\Z; \mu_{\hat\Z}, \sigma_{\hat\Z}^2 I)$, where $\Z$ and $\hat\Z$ denote the random variables of $\z$ and $\hat\z$, respectively. The FID metric is then computed from
\begin{align}
\label{eq:fid}
    {\rm FID}(\Z, \hat\Z) = (\mu_{\Z} - \mu_{\hat\Z})^2 + (\sigma_{\Z} - \sigma_{\hat\Z})^2.
\end{align}
}

For the vessel segmentation, we use a pre-trained model Iter-Net~\cite{9093621}, which was trained on a distinct DRIVE dataset~\cite{drive_website}, CHASE-DB1~\cite{Owen2009MeasuringRV}, and STARE~\cite{845178}. We calculated the \add{IOU} score between the segmentation of the clean images and the segmentation of the comparison images.
\add{
IOU is computed by
\begin{align}
    {\rm IOU}(\x, \hat\x) = \frac{\x \cap \hat\x}{\x \cup \hat\x},
\end{align}
where $\x \cap \hat\x$ is the number of discrete pixels that overlap after segmentation, and $\x \cup \hat\x$ denotes the union of discrete pixels after segmentation.
}

\noindent\textbf{Evaluation by Ophthalmologists.~}
Two board-certified ophthalmologists with over 25 years of experience (E.C. and K.Y.) conducted an evaluation study, comparing the quality of the enhanced images using 6 different methods. For each of the 50 images, the ranking was chosen from 1 to 6 (the lower the better), where an equivalent quality was marked to be the same score.

\add{
The evaluation was made on the following three criteria:
\begin{itemize}
  \item Clarity of the vessel structure
  \item Visibility of the retinal lesion
  \item Overall artifacts (e.g. illumination, noise, etc.)
\end{itemize}
When hard to decide which image was better, the two images were magnified to a region with a rich structure to compare the degree of noise.
}

\section{\add{Results}}
\label{sec:results}

\subsection{Simulation study}
\label{sec:simulation_study}

\begin{table}[!thb]
\setlength{\tabcolsep}{5pt}
\centering
\caption{
Quantitative evaluation of the simulation study \add{on the EyeQ dataset}, under two different forward models. \textbf{Bold}: best, \underline{underline}: second best. 
}
\resizebox{\columnwidth}{!}{
\begin{tabular}{lcccccc}
\toprule
\textbf{Forward model} & \multicolumn{3}{c}{\cite{shen2020modeling}} & \multicolumn{3}{c}{\textbf{Ours}} \\
\cmidrule{2-4}
\cmidrule{5-7}
 & {PSNR$\uparrow$} & {FID$\downarrow$} & {IOU$\uparrow$} & {PSNR$\uparrow$} & {FID$\downarrow$} & {IOU$\uparrow$}\\
\midrule
Degraded & 17.62 & 50.71 & 0.645 & 17.42 & 94.92 & 0.546 \\
CLAHE~\cite{setiawan2013color} & 17.35 & 53.56 & 0.747 & 18.83 & 42.32 & 0.792 \\
DCP-BCP~\cite{he2010single} & 16.31 & 87.79 & 0.603 & 16.31 & 51.71 & 0.575 \\
SCR-Net~\cite{li2022structureconsistent} & \add{17.56} & \add{54.91} & \add{0.635} & \add{22.91} & \add{53.97} & \add{0.886} \\
CycleGAN~\cite{zhu2017unpaired} & 27.22 & \underline{10.13} & 0.727 & 23.25 & \add{\underline{18.99}} & \underline{0.907} \\
PCE-Net~\cite{liu2022pseudo} & \underline{28.63} & 28.94 & \add{\underline{0.762}} & \add{\underline{24.27}} & \add{25.15} & \add{0.756} \\
\add{BlindDPS~\cite{chung2023parallel}} & \add{15.40} & \add{90.63} & \add{0.443} & \add{15.42} & \add{84.70} & \add{0.532} \\
\add{LED~\cite{cheng2023learning}} & \add{17.31} & \add{41.10} & \add{0.619} & \add{17.00} & \add{57.80} & \add{0.606} \\
\midrule
\rowcolor{BrickRed!10}
\textbf{FD3 (Ours)} & \textbf{34.57} & \textbf{8.997} & \add{\textbf{0.805}} & \textbf{28.07} & \textbf{6.406} & \textbf{0.926} \\
\bottomrule
\end{tabular}
}
\label{tab:results_main_eyeq}
\end{table}

\begin{table}[!thb]
\setlength{\tabcolsep}{5pt}
\centering
\add{\caption{
Quantitative evaluation of the simulation study on the FPE dataset, under the proposed forward model. \textbf{Bold}: best, \underline{underline}: second best. 
}}
\resizebox{0.6\columnwidth}{!}{
\begin{tabular}{lcccccc}
\toprule
\textbf{Forward model} & \multicolumn{3}{c}{\textbf{Ours}} \\
\cmidrule{2-4}
\cmidrule{5-7}
 & {PSNR$\uparrow$} & {FID$\downarrow$} & {IOU$\uparrow$}\\
\midrule
Degraded & 18.17 & 76.19 & 0.465 \\
CLAHE~\cite{setiawan2013color} & 19.26 & 23.91 & 0.720 \\
DCP-BCP~\cite{he2010single} & 18.11 & 86.43 & 0.597 \\
SCR-Net~\cite{li2022structureconsistent} & 24.76 & \underline{20.23} & 0.815 \\
CycleGAN~\cite{zhu2017unpaired} & 20.81 & 39.94 & 0.775 \\
PCE-Net~\cite{liu2022pseudo} & \underline{24.96} & 24.70 & \underline{0.826} \\
BlindDPS~\cite{chung2023parallel}  & 20.48 & 81.24 & 0.462 \\
\add{LED~\cite{cheng2023learning}} & 18.28 & 61.63 & 0.523 \\
\midrule
\rowcolor{BrickRed!10}
\textbf{FD3 (Ours)} & \textbf{28.55} & \textbf{13.13} & \textbf{0.831} \\
\bottomrule
\end{tabular}
}
\label{tab:results_main_fpe}
\end{table}

\noindent
We conduct a simulation study using two different forward models. One that is given in \cite{shen2020modeling} with \eqref{eq:forward_conjunction}, and the other with the proposed forward model given with \eqref{eq:proposed_forward}. The goal of this experiment is to showcase that the proposed method, FD3, outperforms previous arts regardless of the forward model used, showing that FD3 can effectively learn the inverse mapping of the given forward model. The quantitative metrics are shown in Tab.~\ref{tab:results_main_eyeq}. Here, we see that FD3 achieves \add{superior} results in all three different types of metrics, including distortion, fidelity, and downstream performance. 

PSNR and FID values cannot be directly compared in the two different settings as the ``ground truth'' images are different. In contrast, IOU values stem from the same ground truth vessel segmentation map. Comparing the IOU values of the forward model used in \cite{shen2020modeling} and the proposed forward model, we see a significant increase in the IOU values in the recent deep learning-based techniques. Consequently, all experiments that follow hereafter use our forward model that leverages CLAHE unless specified otherwise. \add{We observe a similar trend for the FPE dataset in Tab.~\ref{tab:results_main_fpe}.}

In Fig.~\ref{fig:results_simulation_main}, FD3 provides image enhancements of high fidelity and quality that are the closest to the ground truth. In contrast, PCE-Net often generates results that are off in color tone, and CycleGAN often hallucinates artifacts or fails to remove the artifacts completely. 
\add{
BlindDPS often produces severe undesirable artifacts that are nonexistent in the image, and LED does not significantly remove the artifacts from the degraded image, as opposed to CycleGAN, PCE-Net, and FD3, showing its limitations in being used as a stand-alone method. We discuss and compare against LED being used as a postprocessing method in Sec.~\ref{sec:led}.
}
In Fig.~\ref{fig:vessel}, we show that our method excels in downstream vessel segmentation, as the vessels are clearly visible after the enhancement scheme.

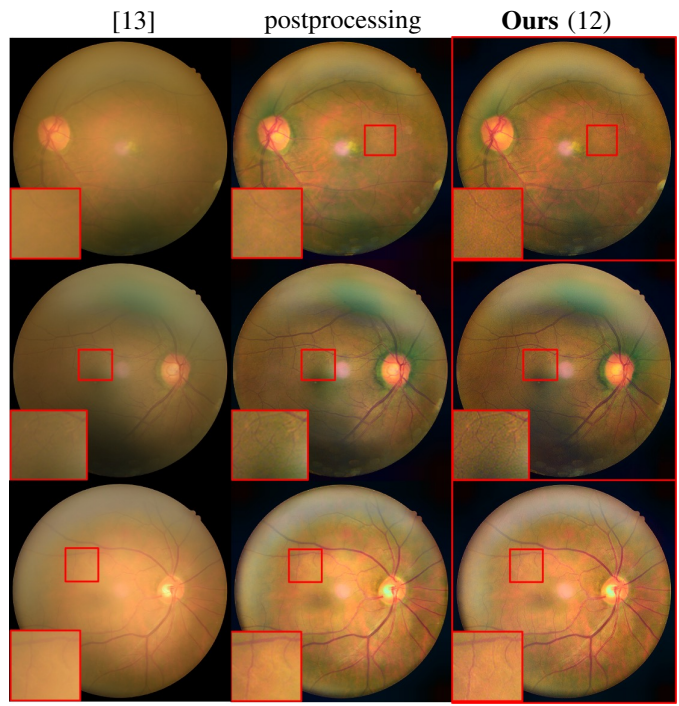
\begin{figure}
    \begin{tikzpicture}
        \node[anchor=south west,inner sep=0] (image) at (0,0) {\includegraphics[width=\columnwidth]{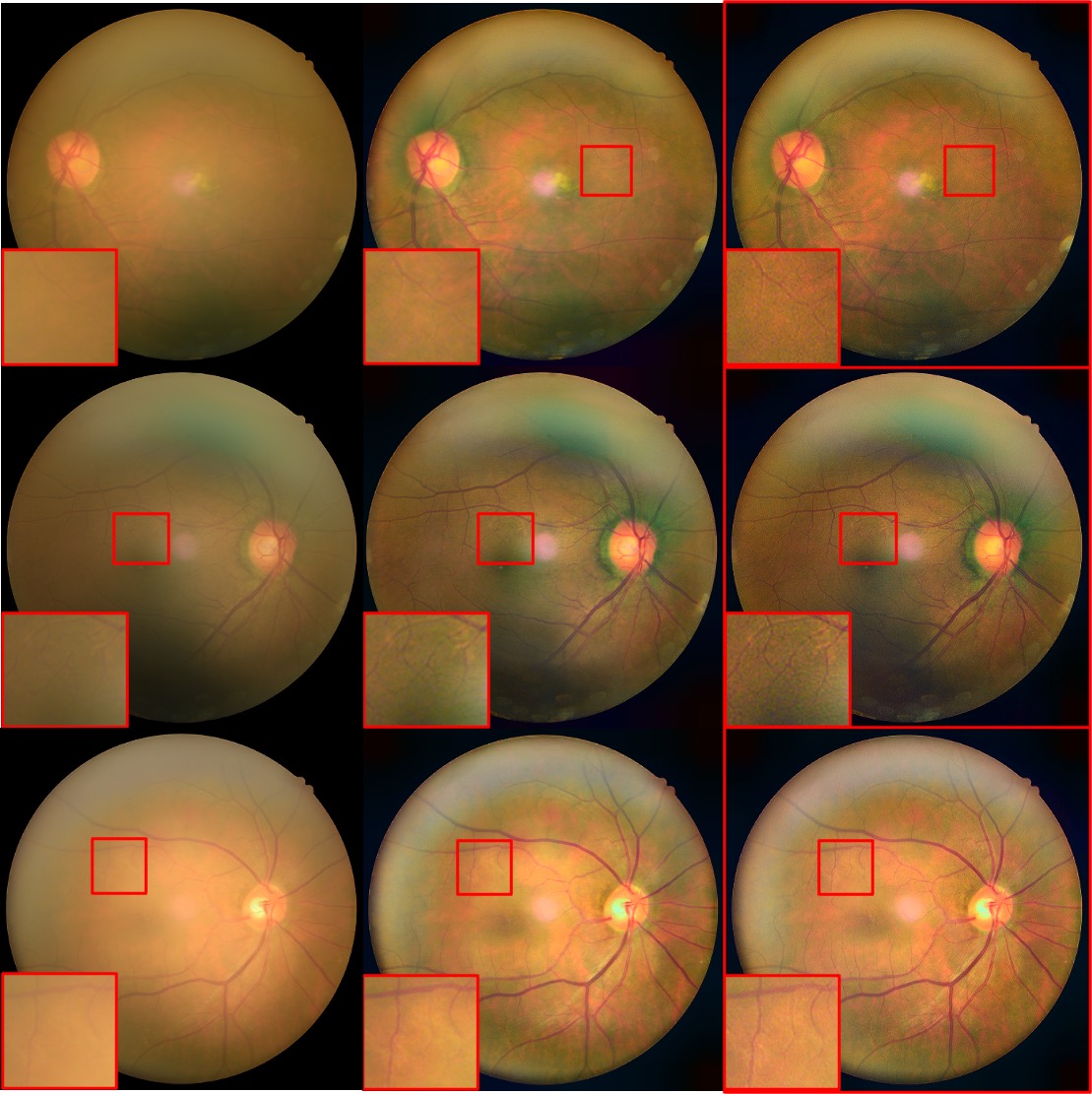}};
        
        \begin{scope}[x={(image.south east)},y={(image.north west)}]
            \node[text width=3cm, align=center] at (0.18,1.02) {\cite{shen2020modeling}};
            \node[text width=3cm, align=center] at (0.5,1.02) {postprocessing};
            \node[text width=3cm, align=center] at (0.82,1.02) {\textbf{Ours} \eqref{eq:proposed_forward}};
        \end{scope}
    \end{tikzpicture}
    \caption{Comparison of in-vivo image enhancement results under different forward models. The model and the inference process are set identically. Column 1: forward model of \cite{shen2020modeling}, column 2: forward model of \cite{shen2020modeling} + CLAHE post-processing, column 3: proposed forward model.}
    \label{fig:comparison_forward_model}
\end{figure}

\begin{table}[!thb]
\setlength{\tabcolsep}{5pt}
\centering
\caption{
Evaluation of 50 in-vivo fundus image enhancements by relative ranking (the lower the better). Average $\pm$ std.
}
\resizebox{0.6\columnwidth}{!}{
\begin{tabular}{lc}
\toprule
& Ranking \\
\midrule
CLAHE~\cite{setiawan2013color} & 2.50 $\pm$ 0.678 \\
DCP-BCP~\cite{he2010single} & 5.40 $\pm$ 0.606 \\
SCR-Net~\cite{li2022structureconsistent} & 5.28 $\pm$ 0.640 \\
CycleGAN~\cite{zhu2017unpaired} & 4.44 $\pm$ 0.837 \\
PCE-Net~\cite{liu2022pseudo} & 2.38 $\pm$ 0.567 \\
\midrule
\rowcolor{BrickRed!10}
\textbf{FD3 (Ours)} & \textbf{1.06} $\pm$ 0.240 \\
\bottomrule
\end{tabular}
}
\label{tab:results_invivo_metric}
\end{table}

\subsection{In vivo study}
\label{sec:in_vivo_study}

In Fig.~\ref{fig:results_invivo_main}, we present a comparison of how various models perform when applied to real ``bad-quality'' images. Our model demonstrated exceptional proficiency in effectively removing haze, resulting in the generation of highly natural-looking eye images. Additionally, our model exhibited superior vessel recognition capabilities compared to other models. Notably, the third row of results stands out as a key highlight. While other models struggled to adequately restore the shadow artifact regions, SCR-Net managed to address this issue but at the expense of retaining vessel details. In contrast, our model not only preserved the eye's shape but also enhanced the visibility of vessels in that specific area.

In Tab.~\ref{tab:results_invivo_metric}, we summarize the evaluation made by the ophthalmologists \add{under the criterion presented in Sec.~\ref{sec:exp}}. We clearly see that FD3 far outperforms the comparison methods. The ophthalmologists both observed that DCP-BCP and SCR-Net sometimes exaggerate the hemorrhage expressed in the images; CycleGAN alleviates the haze artifacts pretty well but often has black-dot artifacts, which could be a serious problem that may lead to misdiagnosis; PCE-Net enhancements are often blurry; CLAHE does not enhance the peripheral parts of the image as well as FD3.

\noindent
\textbf{Choice of the forward model.~}
Recall that one of the main contributions of our work is to devise a better forward model more suited for enhancing the quality of the in-vivo fundus photos. In Fig.~\ref{fig:comparison_forward_model}, we show the superiority of the proposed forward model by comparing it against \cite{shen2020modeling}, and additionally using CLAHE as a post-processing step. It is evident from the figure that only using the forward model of \cite{shen2020modeling} produces unclear and blurry results. Moreover, even if we try to post-process the images through CLAHE, the sharpness, and the microvascular structures are less visible than when we incorporate CLAHE directly into the training process.

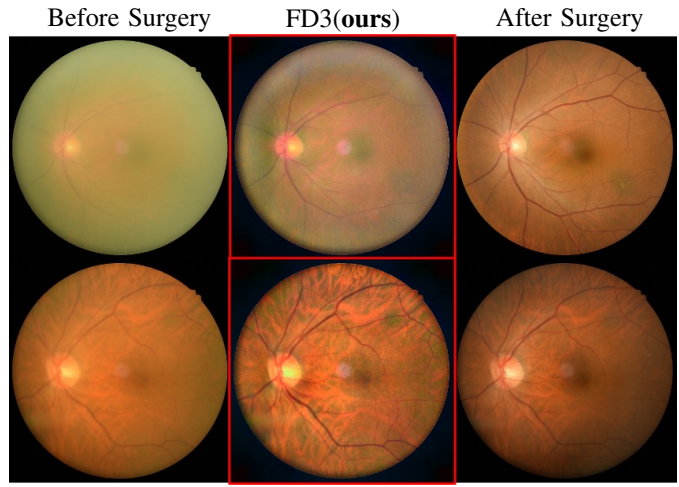
\begin{figure}
    \begin{tikzpicture}
        \node[anchor=south west,inner sep=0] (image) at (0,0) {\includegraphics[width=\columnwidth]{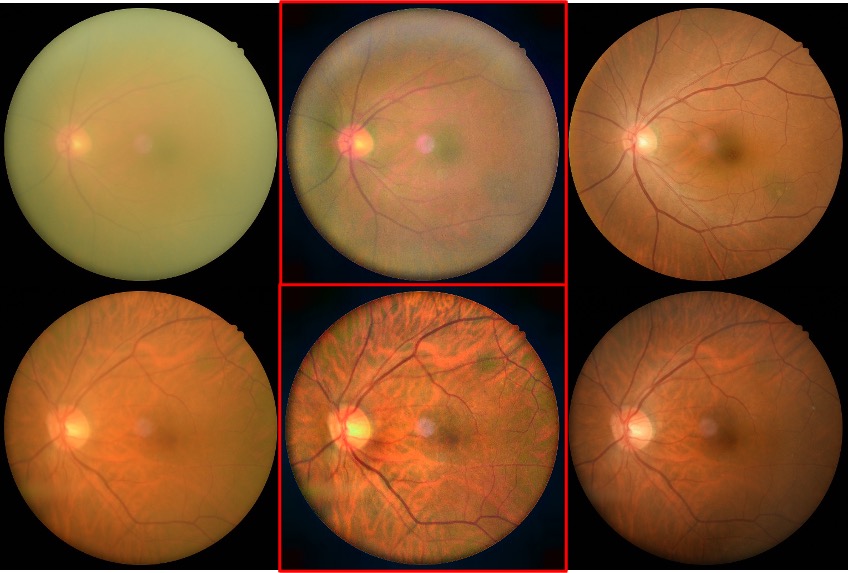}};
        
        \begin{scope}[x={(image.south east)},y={(image.north west)}]
            
            \node[text width=3cm, align=center] at (0.18,1.03) {Before Surgery};
            \node[text width=3cm, align=center] at (0.5,1.03) {FD3(\textbf{ours})};
            \node[text width=3cm, align=center] at (0.84,1.03) {After Surgery};
        \end{scope}
    \end{tikzpicture}
    \caption{Comparison of the fundus photo of patients before and after cataract surgery, and the enhancement result using FD3. After surgery, the removal of the cataract restores clarity to the eye, making the post-cataract eye condition a plausible representation of the ground truth.}
    \label{fig:cataract}
\end{figure}

We further conducted a quantitative evaluation with two ophthalmologists, ranking the relative quality with 50 enhanced in-vivo images (1: better, 2: worse; both marked as 1 if there is no difference in the quality) between using \cite{shen2020modeling} as the forward model with CLAHE postprocessing step and using our forward model. Our method achieved an average ranking of 1.0, whereas \cite{shen2020modeling} + CLAHE postprocessing marked 1.68, meaning that our forward model {\em always} outperformed the counterpart, clearly showing the superiority of our approach.

\noindent
\textbf{Enhancemenet of fundus photos with cataract: A case study.~}
One of the most prominent causes of opaqueness in the bad-quality fundus photos is due to the turbid medium stemming from cataracts. Hence, when taking a fundus photo of a patient before going through the cataract surgery, it would be useful to be able to acquire a high-quality fundus photo with the enhancement that matches the quality of the fundus photo taken after the surgery. To this end, we conducted a study by collecting pre-operative and post-operative fundus photos from the same patient. We compared that to the enhanced photo from the pre-operative fundus image with FD3. In Fig.~\ref{fig:cataract}, we see that for relatively mild degradations, FD3 was capable of providing an enhancement that could fully capture the information that was only available after a fundus photo taken after surgery (row 2). For severe degradations, the effect was less dramatic, yet it was able to improve the quality.

\section{Related works}
\label{sec:related_works}

\noindent
\textbf{Fundus photo enhancement.~}
Numerous efforts have been made to improve the quality of fundus images. One line of approaches involves the combination of  Contrast limited adaptive histogram equalization (CLAHE) and manipulations in the Fourier domain to enhance the clarity of degraded fundus images~\cite{setiawan2013color,jintasuttisak2014color,zhou2017color,article}. Furthermore, to address the issue of the turbidity of fundus images, various techniques have been employed. Dark Channel Prior (DCP) ~\cite{he2010single} and Guided image filtering (GIF)~\cite{6319316} were shown to be effective for removing haze. On the opposite point, bright channel prior (BCP)~\cite{wang2013automatic} was shown to be effective for dealing with under-illumination. 
Structure-preserving guided retinal image filtering (SGRIF) ~\cite{Cheng_2018} has been proposed to restore the fundus images affected by cataracts. However, it is worth noting that the complexity inherent in fundus photography poses a significant challenge when attempting to restore images using these hand-crafted algorithms.

More recently, deep learning-based methods designed to comprehend the characteristics of cataract images have gained popularity. These approaches fall into two categories: degradation modeling-based methods and unpaired image translation. The degradation modeling-based methods aim to understand and rectify image degradation through explicit modeling. \cite{shen2020modeling} proposed the degradation model and developed the fundus enhancement network (Cofe-Net). SCR-Net ~\cite{li2022structureconsistent} tried to maintain the structure consistency by leveraging high-frequency components extracted from synthesized cataract images. Additionally, PCE-Net ~\cite{Liu_2022} extracts multi-level features from Laplacian Pyramid Features to enhance clinically relevant representation, thereby improving the structural information of fundus images and yielding higher-quality cataract images. Nevertheless, these methods have shown limitations in their effectiveness when applied to real clinical data, which presents a more challenging scenario.
On the other hand, unpaired image translation-based methods are trained without any supervision to achieve improved performance on real-world low-quality fundus images. I-SECRET ~\cite{10.1007/978-3-030-87237-3_9} introduced a semi-supervised framework for enhancing fundus images. It includes an unsupervised learning component trained on the unpaired fundus images to enhance its generalization ability. StillGAN ~\cite{ma2021structure} employed an unpaired learning framework that treats low-quality and high-quality images as distinct domains, learning specific enhancement mappings for each. SSGAN-ASP ~\cite{wu2023fundus} introduced the semi-supervised GAN that utilizes the generator to preserve the anatomical structure. However, these models fall short of fully restoring certain fundus characteristics like vessels. Arc-Net ~\cite{li2022annotation} developed an enhancement method learned through unsupervised domain adaptation from the synthesized data. However, this method still struggled to restore low-quality images affected by out-of-distributions (OOD) factors.

\noindent
\textbf{Diffusion model for fundus image enhancement.~}
We are aware of one work that employs diffusion models for fundus image enhancement: LED~\cite{cheng2023learning}.
LED employed a conditional diffusion model to enhance the degraded fundus images. 
However, rather than being a stand-alone approach for image enhancement, LED acts more as a {\em refiner}, that additionally improves the performance of other established methods, which is achieved through an ad-hoc forward-reverse diffusion sampling technique similar to~\cite{chung2022come}.
In contrast to LED~\cite{cheng2023learning}, our model, FD3, adopted a direct diffusion bridge that works as a stand-alone enhancer.

\section{\add{Discussion}}
\label{sec:discussion}

\subsection{\add{Comparison against BlindDPS~\cite{chung2023parallel}}}
\label{sec:blinddps_comparison}

\begin{figure}
    \begin{tikzpicture}
        \node[anchor=south west,inner sep=0] (image) at (0,0) {\includegraphics[width=\columnwidth]{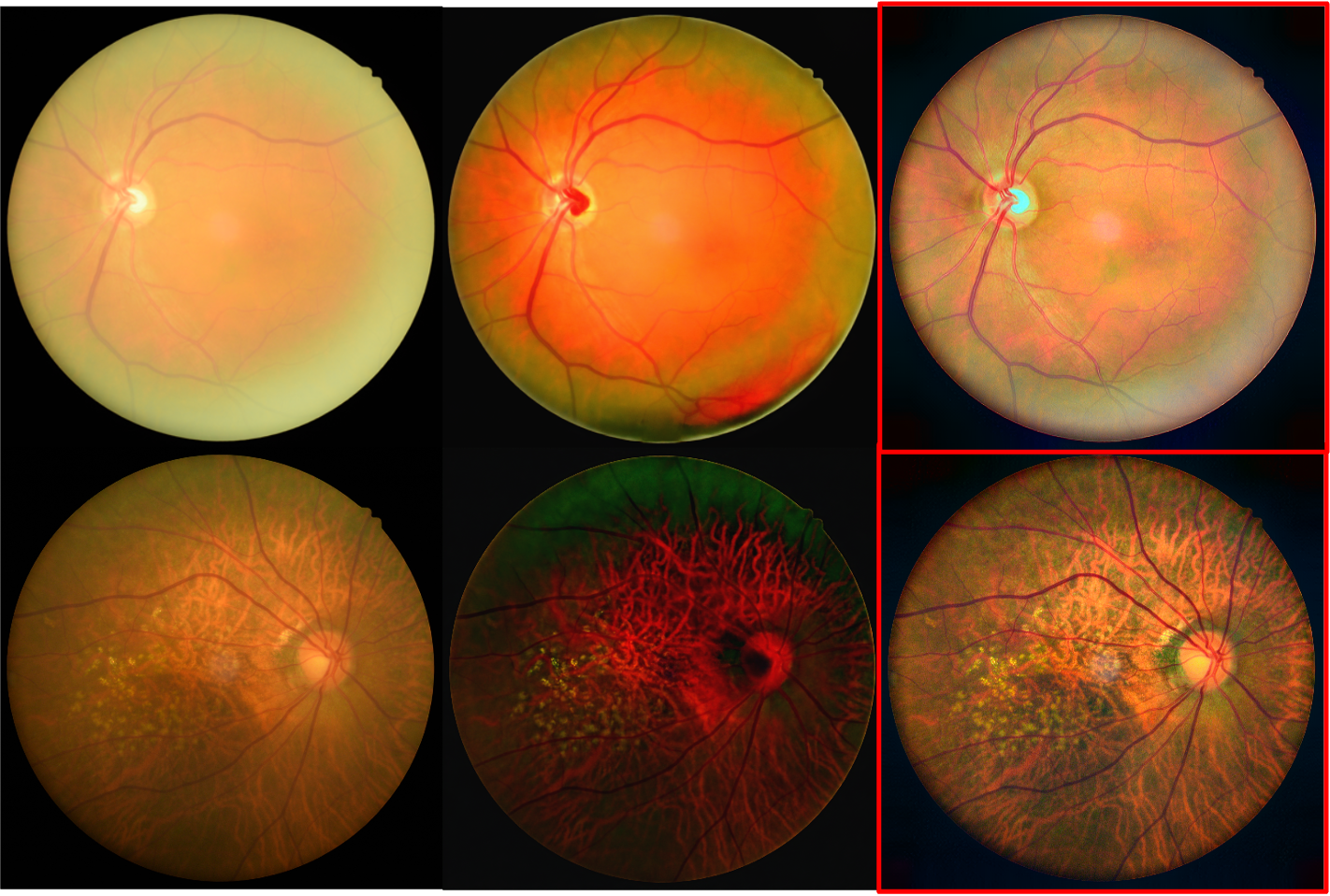}};
        
        \begin{scope}[x={(image.south east)},y={(image.north west)}]
            \node[text width=3cm, align=center] at (0.18,1.03) {Input};
            \node[text width=3cm, align=center] at (0.5,1.03) {BlindDPS};
            \node[text width=3cm, align=center] at (0.84,1.03) {FD3(\textbf{ours})};
        \end{scope}
    \end{tikzpicture}
    \caption{Comparison of FD3 against BlindDPS~\cite{chung2023parallel}.}
    \label{fig:blinddps}
\end{figure}
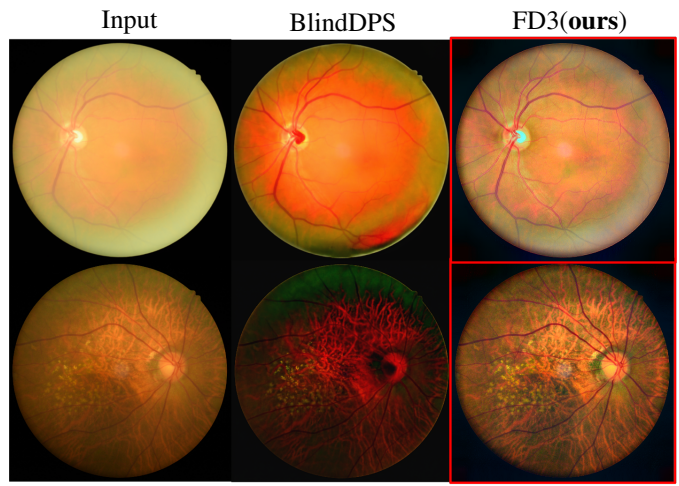

Our model is a direct diffusion bridge that has a more desirable path when solving an inverse problem, as opposed to using a standard diffusion model (\add{See the ``Desirable path'' paragraph of Sec.~\ref{sec:fd3} and recall the difference between \eqref{eq:pf_ode_posterior} and \eqref{eq:fd3_ode}}). Moreover, it has a strong advantage when the forward model is ambiguous, as the inversion of an arbitrary imaging process can be amortized while training the neural network $F_\theta$. To see this in effect, we conduct an experiment where we compare our proposed FD3 against BlindDPS~\cite{chung2023parallel}, which leverages a standard diffusion model that tries to explicitly estimate the parameters of the forward process as well as the underlying ground truth image.

For simplicity, for BlindDPS, we assume that the forward model can be characterized as \eqref{eq:dehazing_forward_model}, similar to what was utilized in \cite{qayyum2020single,qayyum2022single}. With the same neural network architecture that was used for the proposed method, we train two diffusion models that estimate $p(\x)$ and $p(\jb)$, where the transmittance maps $\jb$ were pre-computed using a method in~\cite{he2010single} with high-quality fundus photos. The inference process then follows exactly that of \cite{chung2023parallel} with 1000 DDPM steps. In Fig.~\ref{fig:blinddps}, we compare the results obtained through BlindDPS with the proposed method. We see that the results obtained through BlindDPS are highly unstable, often containing undesirable artifacts.

These results confirm that using a DDB-type approach is much more desirable especially when the underlying forward model is ambiguous, and it is hard to leverage a model-based approach. Furthermore, FD3 yields much more stable results thanks to the diffusion path starting directly from the observed measurement $\y$.


\subsection{\add{Comparison against LED~\cite{cheng2023learning}}}
\label{sec:led}

\begin{table}[!thb]
\setlength{\tabcolsep}{5pt}
\centering
\caption{\add{
Quantitative evaluation of the simulation study of fundus photo enhancement, under with LED or without LED. \textbf{Bold}: best, \underline{underline}: second best. 
}}
\resizebox{\columnwidth}{!}{
\begin{tabular}{lcccccc}
\toprule
\textbf{Forward model} & \multicolumn{3}{c}{\cite{shen2020modeling}} & \multicolumn{3}{c}{\textbf{Ours}} \\
\cmidrule{2-4}
\cmidrule{5-7}
 & {PSNR$\uparrow$} & {FID$\downarrow$} & {IOU$\uparrow$} & {PSNR$\uparrow$} & {FID$\downarrow$} & {IOU$\uparrow$}\\
\midrule
Degraded & 17.62 & 50.71 & 0.645 & 17.42 & 94.92 & 0.546 \\
Degraded+LED~\cite{cheng2023learning} & 17.31 & 41.10 & 0.619 & 17.00 & 57.80 & 0.606 \\
SCR-Net~\cite{li2022structureconsistent} & 17.56 & 54.91 & 0.635 & 22.91 & 53.97 & 0.886 \\
SCR-Net~\cite{li2022structureconsistent}+LED~\cite{cheng2023learning} & 19.78 & 77.36 & 0.742 & 21.56 & 44.72 & 0.877 \\
PCE-Net~\cite{liu2022pseudo} & \underline{28.63} & \underline{28.94} & \underline{0.762} & 24.27 & \underline{25.15} & 0.756 \\
PCE-Net~\cite{liu2022pseudo}+LED~\cite{cheng2023learning} & 25.21 & 34.81 & 0.708 & 23.05 & 43.82 & 0.879 \\
\midrule
\rowcolor{BrickRed!10}
\textbf{FD3 (Ours)} & \textbf{34.57} & \textbf{8.997} & \textbf{0.805} & \textbf{28.07} & \textbf{6.406} & \textbf{0.926} \\
FD3 (Ours)+LED~\cite{cheng2023learning} & 26.62 & 31.33 & 0.720 & \underline{24.91} & 32.86 & \underline{0.888} \\
\bottomrule
\end{tabular}
}
\label{tab:results_led_discussion}
\end{table}

\add{
In Section.~\ref{sec:results}, we compared our method against LED~\cite{cheng2023learning}, which, to the best of our knowledge, is the only existing diffusion model tailored for fundus photo enhancement. In our main comparison, LED was used as an end-to-end enhancer without the use of other enhancement methods together. However, it was pointed out in LED~\cite{cheng2023learning} that it can be used as a postprocessing step after a reconstruction through other methods trained under supervision, e.g. SCR-Net. In Tab.~\ref{tab:results_led_discussion}, we thoroughly compare the results of LED used as a postprocessing step combined with various methods, including FD3. Interestingly, we see that while in some cases, LED improves the metrics by some amount, the effect has high variance, often {\em degrading} the image quality heavily. This can be attributed to the fact that in the training phase, LED was trained as a conditional diffusion model conditioned on the {\em degraded} image, while at inference when used as a postprocessing step, it is conditioned on the {\em restored} image. This may lead to out-of-distribution errors and thereby lead to damaging effects, as seen in Tab.~\ref{tab:results_led_discussion}. In contrast, FD3 stably improves both the perception and the distortion metrics, operating as a stand-alone, end-to-end enhancer.
}

\subsection{\add{Control of perception-distortion tradeoff}}
\label{sec:ablation_study}

\begin{figure}
\centerline{\includegraphics [width=8cm]{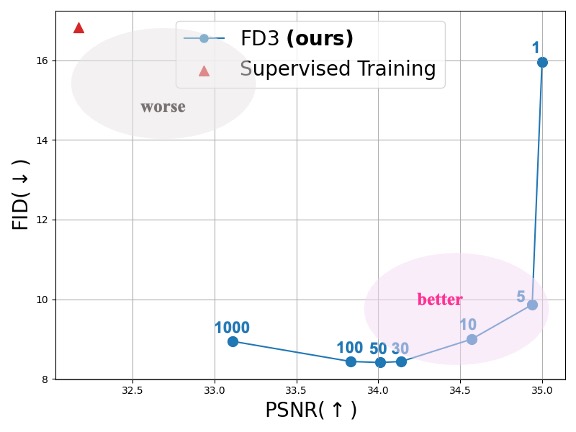}}
\caption{Quantitative metric of FD3 throughout the iteration steps and comparison to the supervised training. Pareto-optimality is achieved in the lower right corner. We choose NFE=10 as it strikes a good balance between PSNR and FID. NFE=10 is chosen over NFE=5 as we opted for better perceptual quality to maximize diagnostic capacity.}
\label{fig:graph}
\end{figure}

Due to the property in \eqref{eq:inference_fd3}, \add{where the choice of timesteps taken is a degree of freedom that only needs to be determined during the inference phase,} we can flexibly control the number of NFEs to achieve a trade-off between perception and distortion. \add{As discussed in Sec.~\ref{sec:fd3}, taking a lower NFE would lead to an estimate closer to the posterior mean, minimizing the distortion. Taking a higher NFE would lead to higher perceptual quality at the cost of moving away from the posterior mean. To verify our hypothesis,}
we plot the trend in Fig.~\ref{fig:graph}. We see that 10 NFE strikes a good balance between the PSNR and the FID score, hence our choice. When opting for better fidelity, one could choose a higher number of NFE with the expense of some distortion.

Furthermore, we conducted a comparative analysis against the simple supervised learning approach, keeping the network architecture and the training process the same, but only using a constant timestep at $t = 1$. Surprisingly, despite the direct inference of targets with the constant timestep, the results were far inferior to FD3. Our hypothesis is that the model gained valuable insights into handling various degrees of degradation when exposed to a random timestep strategy.

\section{Conclusion}
\label{sec:conclusion}

In this work, we presented FD3, a direct diffusion bridge for fundus photo quality enhancement. We devised an effective forward model used for simulation to train our model, which is effective for considering both the local and the global characteristics of the degradation. Our method was robust and capable of producing high-quality restorations, being the first stand-alone diffusion model-based image enhancement method that does not rely on pre-trained restoration models. With extensive experiments in collaboration with board-certified ophthalmologists, we verified that FD3 is exceptionally strong at enhancing the in-vivo fundus photos, achieving \add{exceptional} results.

\section*{References}
\bibliographystyle{ieeetr}
\bibliography{refs}

\begin{thebibliography}{10}

\bibitem{silva2015diabetic}
P.~S. Silva, A.~J.~D. Cruz, M.~G. Ledesma, J.~van Hemert, A.~Radwan, J.~D. Cavallerano, L.~M. Aiello, J.~K. Sun, and L.~P. Aiello, ``Diabetic retinopathy severity and peripheral lesions are associated with nonperfusion on ultrawide field angiography,'' {\em Ophthalmology}, vol.~122, no.~12, pp.~2465--2472, 2015.

\bibitem{issa2013macular}
P.~C. Issa, M.~C. Gillies, E.~Y. Chew, A.~C. Bird, T.~F. Heeren, T.~Peto, F.~G. Holz, and H.~P. Scholl, ``Macular telangiectasia type 2,'' {\em Progress in retinal and eye research}, vol.~34, pp.~49--77, 2013.

\bibitem{niemeijer2005automatic}
M.~Niemeijer, B.~Van~Ginneken, J.~Staal, M.~S. Suttorp-Schulten, and M.~D. Abr{\`a}moff, ``Automatic detection of red lesions in digital color fundus photographs,'' {\em IEEE Transactions on medical imaging}, vol.~24, no.~5, pp.~584--592, 2005.

\bibitem{wong2008natural}
T.~Wong, U.~Chakravarthy, R.~Klein, P.~Mitchell, G.~Zlateva, R.~Buggage, K.~Fahrbach, C.~Probst, and I.~Sledge, ``The natural history and prognosis of neovascular age-related macular degeneration: a systematic review of the literature and meta-analysis,'' {\em Ophthalmology}, vol.~115, no.~1, pp.~116--126, 2008.

\bibitem{philip2005impact}
S.~Philip, L.~Cowie, and J.~Olson, ``The impact of the health technology board for scotland’s grading model on referrals to ophthalmology services,'' {\em British Journal of Ophthalmology}, vol.~89, no.~7, pp.~891--896, 2005.

\bibitem{wong2001retinal}
T.~Y. Wong, R.~Klein, B.~E. Klein, J.~M. Tielsch, L.~Hubbard, and F.~J. Nieto, ``Retinal microvascular abnormalities and their relationship with hypertension, cardiovascular disease, and mortality,'' {\em Survey of ophthalmology}, vol.~46, no.~1, pp.~59--80, 2001.

\bibitem{peli1989restoration}
E.~Peli and T.~Peli, ``Restoration of retinal images obtained through cataracts,'' {\em IEEE transactions on medical imaging}, vol.~8, no.~4, pp.~401--406, 1989.

\bibitem{narasimhan2002vision}
S.~G. Narasimhan and S.~K. Nayar, ``Vision and the atmosphere,'' {\em International journal of computer vision}, vol.~48, pp.~233--254, 2002.

\bibitem{fattal2008single}
R.~Fattal, ``Single image dehazing,'' {\em ACM transactions on graphics (TOG)}, vol.~27, no.~3, pp.~1--9, 2008.

\bibitem{setiawan2013color}
A.~W. Setiawan, T.~R. Mengko, O.~S. Santoso, and A.~B. Suksmono, ``Color retinal image enhancement using clahe,'' in {\em International conference on ICT for smart society}, pp.~1--3, IEEE, 2013.

\bibitem{jintasuttisak2014color}
T.~Jintasuttisak and S.~Intajag, ``Color retinal image enhancement by rayleigh contrast-limited adaptive histogram equalization,'' in {\em 2014 14th international conference on control, automation and systems (ICCAS 2014)}, pp.~692--697, IEEE, 2014.

\bibitem{zhou2017color}
M.~Zhou, K.~Jin, S.~Wang, J.~Ye, and D.~Qian, ``Color retinal image enhancement based on luminosity and contrast adjustment,'' {\em IEEE Transactions on Biomedical engineering}, vol.~65, no.~3, pp.~521--527, 2017.

\bibitem{shen2020modeling}
Z.~Shen, H.~Fu, J.~Shen, and L.~Shao, ``Modeling and enhancing low-quality retinal fundus images,'' {\em IEEE transactions on medical imaging}, vol.~40, no.~3, pp.~996--1006, 2020.

\bibitem{qayyum2020single}
A.~Qayyum, W.~Sultani, F.~Shamshad, J.~Qadir, and R.~Tufail, ``Single-shot retinal image enhancement using deep image priors,'' in {\em Medical Image Computing and Computer Assisted Intervention--MICCAI 2020: 23rd International Conference, Lima, Peru, October 4--8, 2020, Proceedings, Part V 23}, pp.~636--646, Springer, 2020.

\bibitem{qayyum2022single}
A.~Qayyum, W.~Sultani, F.~Shamshad, R.~Tufail, and J.~Qadir, ``Single-shot retinal image enhancement using untrained and pretrained neural networks priors integrated with analytical image priors,'' {\em Computers in Biology and Medicine}, vol.~148, p.~105879, 2022.

\bibitem{luo2020dehaze}
Y.~Luo, K.~Chen, L.~Liu, J.~Liu, J.~Mao, G.~Ke, and M.~Sun, ``Dehaze of cataractous retinal images using an unpaired generative adversarial network,'' {\em IEEE Journal of Biomedical and Health Informatics}, vol.~24, no.~12, pp.~3374--3383, 2020.

\bibitem{he2010single}
K.~He, J.~Sun, and X.~Tang, ``Single image haze removal using dark channel prior,'' {\em IEEE transactions on pattern analysis and machine intelligence}, vol.~33, no.~12, pp.~2341--2353, 2010.

\bibitem{ulyanov2018deep}
D.~Ulyanov, A.~Vedaldi, and V.~Lempitsky, ``Deep image prior,'' in {\em Proceedings of the IEEE conference on computer vision and pattern recognition}, pp.~9446--9454, 2018.

\bibitem{sohl2015deep}
J.~Sohl-Dickstein, E.~Weiss, N.~Maheswaranathan, and S.~Ganguli, ``Deep unsupervised learning using nonequilibrium thermodynamics,'' in {\em International Conference on Machine Learning}, pp.~2256--2265, PMLR, 2015.

\bibitem{ho2020denoising}
J.~Ho, A.~Jain, and P.~Abbeel, ``Denoising diffusion probabilistic models,'' {\em Advances in Neural Information Processing Systems}, vol.~33, pp.~6840--6851, 2020.

\bibitem{song2020score}
Y.~Song, J.~Sohl{-}Dickstein, D.~P. Kingma, A.~Kumar, S.~Ermon, and B.~Poole, ``Score-based generative modeling through stochastic differential equations,'' in {\em 9th International Conference on Learning Representations, {ICLR}}, 2021.

\bibitem{karras2022elucidating}
T.~Karras, M.~Aittala, T.~Aila, and S.~Laine, ``Elucidating the design space of diffusion-based generative models,'' in {\em Proc. NeurIPS}, 2022.

\bibitem{efron2011tweedie}
B.~Efron, ``Tweedie’s formula and selection bias,'' {\em Journal of the American Statistical Association}, vol.~106, no.~496, pp.~1602--1614, 2011.

\bibitem{vincent2011connection}
P.~Vincent, ``A connection between score matching and denoising autoencoders,'' {\em Neural computation}, vol.~23, no.~7, pp.~1661--1674, 2011.

\bibitem{kadkhodaie2021stochastic}
Z.~Kadkhodaie and E.~P. Simoncelli, ``Stochastic solutions for linear inverse problems using the prior implicit in a denoiser,'' in {\em Advances in Neural Information Processing Systems} (A.~Beygelzimer, Y.~Dauphin, P.~Liang, and J.~W. Vaughan, eds.), 2021.

\bibitem{kawar2022denoising}
B.~Kawar, M.~Elad, S.~Ermon, and J.~Song, ``Denoising diffusion restoration models,'' in {\em Advances in Neural Information Processing Systems} (A.~H. Oh, A.~Agarwal, D.~Belgrave, and K.~Cho, eds.), 2022.

\bibitem{chung2023diffusion}
H.~Chung, J.~Kim, M.~T. Mccann, M.~L. Klasky, and J.~C. Ye, ``Diffusion posterior sampling for general noisy inverse problems,'' in {\em International Conference on Learning Representations}, 2023.

\bibitem{chung2023parallel}
H.~Chung, J.~Kim, S.~Kim, and J.~C. Ye, ``Parallel diffusion models of operator and image for blind inverse problems,'' {\em IEEE/CVF Conference on Computer Vision and Pattern Recognition}, 2023.

\bibitem{murata2023gibbsddrm}
N.~Murata, K.~Saito, C.-H. Lai, Y.~Takida, T.~Uesaka, Y.~Mitsufuji, and S.~Ermon, ``Gibbsddrm: A partially collapsed gibbs sampler for solving blind inverse problems with denoising diffusion restoration,'' in {\em International conference on machine learning}, PMLR, 2023.

\bibitem{chung2023direct}
H.~Chung, J.~Kim, and J.~C. Ye, ``Direct diffusion bridge using data consistency for inverse problems,'' {\em Advances in Neural Information Processing Systems}, 2023.

\bibitem{delbracio2023inversion}
M.~Delbracio and P.~Milanfar, ``Inversion by direct iteration: An alternative to denoising diffusion for image restoration,'' {\em Transactions on Machine Learning Research}, 2023.
\newblock Featured Certification.

\bibitem{liu2023i2sb}
G.-H. Liu, A.~Vahdat, D.-A. Huang, E.~A. Theodorou, W.~Nie, and A.~Anandkumar, ``{I$^2$SB: Image-to-Image Schr{\"o}dinger Bridge},'' in {\em International conference on machine learning}, PMLR, 2023.

\bibitem{Fu_2019}
H.~Fu, B.~Wang, J.~Shen, S.~Cui, Y.~Xu, J.~Liu, and L.~Shao, ``Evaluation of retinal image quality assessment networks in different color-spaces,'' in {\em Lecture Notes in Computer Science}, pp.~48--56, Springer International Publishing, 2019.

\bibitem{liu2023flow}
X.~Liu, C.~Gong, and qiang liu, ``Flow straight and fast: Learning to generate and transfer data with rectified flow,'' in {\em The Eleventh International Conference on Learning Representations}, 2023.

\bibitem{blau2018perception}
Y.~Blau and T.~Michaeli, ``The perception-distortion tradeoff,'' in {\em Proceedings of the IEEE conference on computer vision and pattern recognition}, pp.~6228--6237, 2018.

\bibitem{zhu2017unpaired}
J.-Y. Zhu, T.~Park, P.~Isola, and A.~A. Efros, ``Unpaired image-to-image translation using cycle-consistent adversarial networks,'' in {\em Proceedings of the IEEE international conference on computer vision}, pp.~2223--2232, 2017.

\bibitem{Liu_2022}
H.~Liu, H.~Li, H.~Fu, R.~Xiao, Y.~Gao, Y.~Hu, and J.~Liu, ``Degradation-invariant enhancement of fundus images via pyramid constraint network,'' in {\em Lecture Notes in Computer Science}, pp.~507--516, Springer Nature Switzerland, 2022.

\bibitem{cheng2023learning}
P.~Cheng, L.~Lin, Y.~Huang, H.~He, W.~Luo, and X.~Tang, ``Learning enhancement from degradation: A diffusion model for fundus image enhancement,'' 2023.

\bibitem{9093621}
L.~Li, M.~Verma, Y.~Nakashima, H.~Nagahara, and R.~Kawasaki, ``Iternet: Retinal image segmentation utilizing structural redundancy in vessel networks,'' in {\em 2020 IEEE Winter Conference on Applications of Computer Vision (WACV)}, pp.~3645--3654, 2020.

\bibitem{li2022structureconsistent}
H.~Li, H.~Liu, H.~Fu, H.~Shu, Y.~Zhao, X.~Luo, Y.~Hu, and J.~Liu, ``Structure-consistent restoration network for cataract fundus image enhancement,'' 2022.

\bibitem{dhariwal2021diffusion}
P.~Dhariwal and A.~Q. Nichol, ``Diffusion models beat {GAN}s on image synthesis,'' in {\em Advances in Neural Information Processing Systems} (A.~Beygelzimer, Y.~Dauphin, P.~Liang, and J.~W. Vaughan, eds.), 2021.

\bibitem{wang2013automatic}
Y.~Wang, S.~Zhuo, D.~Tao, J.~Bu, and N.~Li, ``Automatic local exposure correction using bright channel prior for under-exposed images,'' {\em Signal processing}, vol.~93, no.~11, pp.~3227--3238, 2013.

\bibitem{liu2022pseudo}
L.~Liu, Y.~Ren, Z.~Lin, and Z.~Zhao, ``Pseudo numerical methods for diffusion models on manifolds,'' in {\em International Conference on Learning Representations}, 2022.

\bibitem{szegedy2015going}
C.~Szegedy, W.~Liu, Y.~Jia, P.~Sermanet, S.~Reed, D.~Anguelov, D.~Erhan, V.~Vanhoucke, and A.~Rabinovich, ``Going deeper with convolutions,'' in {\em Proceedings of the IEEE conference on computer vision and pattern recognition}, pp.~1--9, 2015.

\bibitem{drive_website}
DRIVE, ``Digital retinal images for vessel extraction (drive).'' \url{https://drive.grand-challenge.org/}, 2019.

\bibitem{Owen2009MeasuringRV}
C.~G. Owen, A.~R. Rudnicka, R.~Mullen, S.~A. Barman, D.~N. Monekosso, P.~H. Whincup, J.~Ng, and C.~Paterson, ``Measuring retinal vessel tortuosity in 10-year-old children: validation of the computer-assisted image analysis of the retina (caiar) program.,'' {\em Investigative ophthalmology \& visual science}, vol.~50 5, pp.~2004--10, 2009.

\bibitem{845178}
A.~Hoover, V.~Kouznetsova, and M.~Goldbaum, ``Locating blood vessels in retinal images by piecewise threshold probing of a matched filter response,'' {\em IEEE Transactions on Medical Imaging}, vol.~19, no.~3, pp.~203--210, 2000.

\bibitem{article}
A.~Mitra, S.~Roy, S.~Roy, and S.~Setua, ``Enhancement and restoration of non-uniform illuminated fundus image of retina obtained through thin layer of cataract,'' {\em Computer Methods and Programs in Biomedicine}, vol.~156, 01 2018.

\bibitem{6319316}
K.~He, J.~Sun, and X.~Tang, ``Guided image filtering,'' {\em IEEE Transactions on Pattern Analysis and Machine Intelligence}, vol.~35, no.~6, pp.~1397--1409, 2013.

\bibitem{Cheng_2018}
J.~Cheng, Z.~Li, Z.~Gu, H.~Fu, D.~W.~K. Wong, and J.~Liu, ``Structure-preserving guided retinal image filtering and its application for optic disk analysis,'' {\em {IEEE} Transactions on Medical Imaging}, vol.~37, pp.~2536--2546, nov 2018.

\bibitem{10.1007/978-3-030-87237-3_9}
P.~Cheng, L.~Lin, Y.~Huang, J.~Lyu, and X.~Tang, ``I-secret: Importance-guided fundus image enhancement via semi-supervised contrastive constraining,'' in {\em Medical Image Computing and Computer Assisted Intervention -- MICCAI 2021} (M.~de~Bruijne, P.~C. Cattin, S.~Cotin, N.~Padoy, S.~Speidel, Y.~Zheng, and C.~Essert, eds.), (Cham), pp.~87--96, Springer International Publishing, 2021.

\bibitem{ma2021structure}
Y.~Ma, J.~Liu, Y.~Liu, H.~Fu, Y.~Hu, J.~Cheng, H.~Qi, Y.~Wu, J.~Zhang, and Y.~Zhao, ``Structure and illumination constrained gan for medical image enhancement,'' {\em IEEE Transactions on Medical Imaging}, vol.~40, no.~12, pp.~3955--3967, 2021.

\bibitem{wu2023fundus}
H.-T. Wu, X.~Cao, Y.~Gao, K.~Zheng, J.~Huang, J.~Hu, and Z.~Tian, ``Fundus image enhancement via semi-supervised gan and anatomical structure preservation,'' {\em IEEE Transactions on Emerging Topics in Computational Intelligence}, 2023.

\bibitem{li2022annotation}
H.~Li, H.~Liu, Y.~Hu, H.~Fu, Y.~Zhao, H.~Miao, and J.~Liu, ``An annotation-free restoration network for cataractous fundus images,'' {\em IEEE Transactions on Medical Imaging}, vol.~41, no.~7, pp.~1699--1710, 2022.

\bibitem{chung2022come}
H.~Chung, B.~Sim, and J.~C. Ye, ``{Come-Closer-Diffuse-Faster: Accelerating Conditional Diffusion Models for Inverse Problems through Stochastic Contraction},'' in {\em Proceedings of the IEEE/CVF Conference on Computer Vision and Pattern Recognition}, 2022.

\end{thebibliography}

\end{document}